\documentclass[a4paper,11pt]{article}
\usepackage{jheppub}
\usepackage[utf8x]{inputenc}
\usepackage[T1]{fontenc}
\usepackage{braket}
\usepackage{physics}
\usepackage{color}
\usepackage{amsfonts}
\usepackage{amssymb}
\usepackage{amsmath}
\usepackage[normalem]{ulem}
\usepackage[font={footnotesize}]{caption}
\usepackage{graphicx}
\usepackage{epstopdf}
\usepackage{enumitem}
\usepackage{subcaption}
\usepackage{comment}
\usepackage{mathtools}
\usepackage{ragged2e}
\usepackage{hyperref}
\hypersetup{
	colorlinks=true,
	linkcolor=blue,
	filecolor=cyan, 
	urlcolor=blue,
	citecolor=red,
} 
\usepackage[titletoc,toc,title]{appendix}
\numberwithin{equation}{section}
\usepackage[nameinlink]{cleveref}
\usepackage{float}
\makeatletter
\newcommand{\vast}{\bBigg@{3}}
\newcommand{\Vast}{\bBigg@{5}}
\makeatother

\newcommand{\TTbar}{\text{T}\bar{\text{T}}}

\newcommand{\zbar}{\raisebox{0.2ex}{--}\kern-0.6em Z}

\def\CA{{\cal A}}

\def\CC{{\cal C}}
\def\CD{{\cal D}}

\def\CF{{\cal F}}

\def\CL{{\cal L}}

\def\CM{{\cal M}}

\def\CO{{\cal O}}

\def\CS{{\cal S}}

\def\BR{\mathbb{R}}

\def\d{\textrm{d}}
\def\del{\partial}

\def\TTbar{\textrm{T}\overline{\textrm{T}}}

\begin{document}
	\title{Butterfly effect and $\TTbar$-deformation}

	\author[\text{a}]{Debarshi Basu, Ashish Chandra, Qiang Wen}

	\affiliation[\text{a}]{
		Shing-Tung Yau Center and School of Physics, Southeast University,\\
		Nanjing 210096, China\\
	}
	%	\affiliation[\bar{\text{T}}]{
		%		Department of Physics,\\
		%		Indian Institute of Technology,\\ 
		%		Kanpur 208 016, India
		%	}
	
	\emailAdd{debarshi.128@gmail.com, achadrahep@gmail.com, wenqiang@seu.edu.cn}

	\abstract{These notes present a comprehensive analysis of shockwave geometries in holographic settings, focusing on $\TTbar$-deformed BTZ black holes and their extensions. By constructing deformed metrics and employing Kruskal coordinates, we examine out-of-time-ordered correlators (OTOCs) as probes of quantum chaos. We also study localized shockwave solutions and analyze their backreaction, highlighting regimes in which the Mezei-Stanford bound on the butterfly velocity is potentially violated. The results obtained via shockwave methods are corroborated with recent developments in pole-skipping phenomena and the entanglement wedge approach, demonstrating consistency among distinct probes of chaos in holographic theories.% \cite{Maldacena:2015waa, Shenker:2013pqa, Mezei:2019dfv, Blake:2018leo}.
	}

	\maketitle
	
	\flushbottom
	\section{Introduction}
	The study of quantum chaos has emerged as a cornerstone of modern theoretical physics, offering profound insights into the dynamics of quantum systems, particularly those governed by strong interactions. In this context, black holes serve as natural objects for exploring quantum chaos, as their near-horizon geometries encode the scrambling of quantum information in dual field theories. Interestingly, various characteristic methods of chaos diagnostics have been proposed in the literature, such as out-of-time-ordered correlators (OTOCs), pole skipping and entanglement wedge method \cite{Dolan:2000ut, Fitzpatrick:2014vua, Roberts:2014ifa, Grozdanov:2017ajz, Mezei:2016wfz}. 
	\begin{itemize}

\item The OTOCs shows an exponential behavior at late time caused due to small perturbations at early time. In particular, this is precisely captured by the growth of four point correlator over time, resulting a characteristic exponential behavior described by the Lyapunov exponent. In chaotic systems, OTOCs typically exhibit exponential growth, approximated as 
	\begin{align}
		\langle V(t) W(0) V(t) W(0) \rangle \approx 1 - \epsilon\, e^{\lambda_L (t-t_*)}\,,
	\end{align}
	where \(\lambda_L\) is the Lyapunov exponent and \(t_*\) is described as the scrambling time, which indicates how quickly information spreads across the quantum system~\cite{Sekino:2008he, Shenker:2013pqa}.
	
	In the context of AdS/CFT correspondence \cite{Maldacena:1997re,Gubser:1998bc}, eternal black holes are dual to thermofield double states of two coupled conformal field theories (CFTs) providing a natural framework to investigate quantum chaos. In this regards, the perturbations at early time applied on one side of the thermofield double state grow exponentially in time. In holographic settings, the OTOC takes the form\footnote{See also \cite{Xu:2018dfp, Gharibyan:2018fax, Steinberg:2019uqb, Gu:2021xaj, Poojary:2018esz, Banerjee:2018twd, Malvimat:2021itk, Das:2022jrr, Das:2021qsd, Das:2019tga,Fan:2016ean,Biswas:2024mlq,Baishya:2024sym,Banerjee:2022ime,Xu:2022vko,Das:2022pez,Hashimoto:2020xfr,Khetrapal:2022dzy,Perlmutter:2016pkf}, for the further studies of OTOCs in various systems. } 
	\begin{align}
		\mathcal{C}(t, x) \sim \exp \left[\lambda_L\left(t - t_* - \frac{|x|}{v_B}\right)\right], 
	\end{align}
	where $v_B$ and $t_\star$ are characteristic speed and time scales referred to as the butterfly velocity and the scrambling time.
	Here, the blueshift experienced by infalling quanta effectively converts a small initial perturbation into a shockwave in the bulk \cite{Roberts:2014ifa, Shenker:2013pqa}. Furthermore, the authors of  \cite{Roberts:2014isa} also extended corresponding analysis for the dual OTOCs in localized shockwave configurations where it involve investigating corresponding RT surface discussed in \cite{Ryu:2006bv,Ryu:2006ef,Hubeny:2007xt}. Interestingly, the OTOCs in the bulk dual geometries depends on chaos parameters, such as the Lyapunov exponent and the butterfly velocity. In addition, it was proposed that these chaos parameters follow certain fundamental bounds. The Maldacena-Shenker-Stanford (MSS) bound limits the Lyapunov exponent to $\lambda_L \leq \frac{2\pi}{\beta}$, where $\beta$ is the inverse temperature \cite{Maldacena:2015waa}, while the Mezei-Stanford bound restricts the butterfly velocity to \cite{Mezei:2016wfz}
	\begin{align}
		v_B \leq v_B^\textrm{Sch}=\sqrt{\frac{d}{2(d-1)}}
	\end{align}
	for theories dual to two-derivative gravity. 
	
	\item 	The chaotic nature can also be captured by the properties of the retarded energy-density Green's function at the so called \textit{pole-skipping} points, which are holographically related to the near horizon properties of metric perturbations in the bulk \cite{Grozdanov:2017ajz,Blake:2018leo, Blake:2017ris, Grozdanov:2018kkt, Blake:2019otz}. Pole-skipping occurs when lines of zeros and poles intersect. The frequency and momentum values where this happens are related to measures of chaos in the following way,  %Blake2018
	\begin{align}
		\omega_\star=i\lambda_L~~,~~k_\star=i\frac{\lambda_L}{v_B}\,.\label{Pole-skipping-point}
	\end{align}
	
\item 	The entanglement wedge method, on the other hand, offers a geometric perspective by analyzing the minimal boundary region containing a falling particle, providing a direct measure of \(v_B\)~\cite{Mezei:2016wfz}. It links quantum states on the boundary to gravitational dynamics in the bulk, where a local perturbation, modeled as a particle falling toward a black hole horizon, causes information to delocalize over time. %Qi2019

		\end{itemize}
	On the other hand, the $\TTbar$ deformation introduces a novel, exactly solvable irrelevant double trace deformations in CFT$_2$s that affects both the ultraviolet and infrared behavior of the theories. This deformation is characterized by a term proportional to the determinant of the stress-energy tensor and governed by a parameter $\mu$. It retains modular invariance \cite{Cardy:2018sdv,Dubovsky:2018bmo,Datta:2018thy,Aharony:2018bad}, and also preserves the integrability of the original theory involving infinite set of conserved charges which provides exact computations of critical quantities such as the energy spectrum, partition function, scattering S-matrix and entanglement and correlation properties \cite{Zamolodchikov:2004ce,Cavaglia:2016oda,Smirnov:2016lqw,He:2019vzf,He:2020cxp,He:2022ryk,He:2023wko,Tian:2023fgf,Tian:2024vln,He:2023obo,He:2023xnb,He:2024pbp}. In the literature, various holographic dual avatars for the $\TTbar$-deformed CFTs have been proposed, including CFT coupled to topological gravity \cite{Dubovsky:2017cnj,Dubovsky:2018bmo}, non-critical string theory \cite{Callebaut:2019omt,Tolley:2019nmm}, AdS$_3$ gravity at a finite radial cut-off \cite{McGough:2016lol,Kraus:2018xrn,Taylor:2018xcy,Kraus:2022mnu} \footnote{See also,  \cite{Asrat:2017tzd,Shyam:2017znq,Cottrell:2018skz,Hartman:2018tkw,Shyam:2018sro,Caputa:2019pam,Lewkowycz:2019xse,Giveon:2017myj,Chang:2024voo} for further investigation in this direction.}, the glue-on AdS holography \cite{Apolo:2023vnm,Apolo:2023ckr}, the mixed boundary conditions prescription in \cite{Guica:2019nzm}, as well as the dual gravity theory \cite{Hirano:2020nwq,Hirano:2020ppu} of the random boundary geometry proposed in \cite{Cardy:2018sdv}. In this article, we will mostly focus on the mixed boundary conditions prescription, wherein the  holographic dual of a $\TTbar$ deformed CFT$_2$ corresponds to AdS$_3$ gravity with mixed non-linear boundary conditions at the asymptotic boundary \cite{Guica:2019nzm}, in contrast to the usual Dirichlet boundary conditions. This prescription is similar to the  cut-off prescription in \cite{McGough:2016lol} in a sense that imposing Dirichlet boundary conditions at a finite radial cutoff in the bulk is effectively equivalent to mixed boundary conditions at the asymptotics, on-shell \cite{Faulkner:2010jy, Heemskerk:2010hk, Balasubramanian:2012hb}. For a pedagogical review of the historical developments and more recent aspects of $\TTbar$ deformation and holography, the readers are directed to the reviews \cite{Jiang:2019epa,He:2025ppz}.
	
%	Moreover, the authors in \cite{McGough:2016lol} proposed a holographic dual of such $\TTbar$ deformed CFTs. The dual bulk theory in this context remains AdS in nature but is now characterized by a Dirichlet holographic screen moved to a finite radial position. Later, an alternative description of the bulk dual was proposed in \cite{Guica:2019nzm} which involves mixed boundary conditions at the usual AdS asymptotic boundary. These proposals have similarity 	

Interestingly, the nature of quantum chaos in the case of $\TTbar$ deformed theories was studied in earlier literature \cite{He:2019vzf} where it was demonstrated that the OTOC in these field theories saturate the MSS bound and exhibit exponential structure at late times. In this work, we explore the impact of $\TTbar$ deformations on holographic quantum chaos. In the holographic context provided in \cite{Guica:2019nzm}, this deformation corresponds to a modified bulk geometry with the deformed field theory located at the asymptotic boundary. This geometric reinterpretation allows one to investigate the effects of $\TTbar$ deformations on the BTZ black hole, dual to a deformed CFT$_2$ at finite temperature.  In this regard, we obtain the OTOC through the computation of geodesic length in the non-rotating deformed BTZ black hole geometry in Kruskal coordinates. We follow corresponding analysis for both spherically symmetric and localized shockwave scenarios \cite{Shenker:2013pqa,Roberts:2014isa}. In this context, we observe a potential violation of the Mezei-Stanford bound \cite{Mezei:2016wfz} on the butterfly velocity. In undeformed holographic theories, this bound ensures that chaos spreads at a speed consistent with causality in the bulk. However, the $\TTbar$ deformation introduces non-local effects that may challenge these expectations, particularly for negative values of the deformation parameter. To probe this, we derive the butterfly velocity using three independent methods: the shockwave approach, pole-skipping phenomena, and the entanglement wedge method.  For rotating BTZ black holes, we extend our analysis to account for the conserved angular momentum. Here, we focus primarily on pole-skipping and shockwave methods. The rotating case allows to explore how the chemical potential for angular momentum modifies chaos parameters, particularly in near extremal limits where the inner and outer horizons coincide. We observe that the Lyapunov exponent consistently saturates the MSS bound across deformed and undeformed scenarios. In contrast, the butterfly velocity exhibits striking behavior: for negative deformation parameters, it can exceed the Mezei-Stanford bound, while near the Hagedorn bound it approaches zero.
	%We also analyze the holographic mutual information in this shockwave geometry.
	
	In what follows, we outline the structure of our paper. In \cref{sec-2}, we present	a detailed review of the construction of $\TTbar$ deformed BTZ black holes and the derivation of the deformed metric in Kruskal coordinates. The \cref{sec-3} is devoted to the derivation of shockwave solutions in these deformed non-rotating BTZ geometries. We explicitly compute the blueshift factors and analyze	OTOCs in the corresponding geometry. In \cref{sec-4}, we carry out a detailed computation of OTOC in the deformed rotating BTZ background, extracting the modified Lyapunov exponent and butterfly velocity. Finally, in \cref{sec-5}, we discuss our results for holographic OTOCs, pole-skipping and the dynamics of quantum information scrambling. We conclude with a summary of our findings and a discussion of possible future directions.

	\section{A brief review on $\TTbar$ deformation and holography}\label{sec-2}
	The $\TTbar$ deformation is a universal irrelevant deformation of a two-dimensional quantum field theory, generated by the determinant of the instantaneous stress tensor, and is governed by the flow equation \cite{Zamolodchikov:2004ce,Cavaglia:2016oda,McGough:2016lol,Guica:2019nzm}:
	\begin{align}
		\del_\mu \CS^{[\mu]}=-\frac{1}{2}\int \d^2x\sqrt{\gamma^{[\mu]}}\left(\gamma_{ab}T^{ac}T^b_{\,\,c}-T^2\right)^{[\mu]}\,,\label{TTbar-defn}
	\end{align}
	where the superscript $[\mu]$ denotes that the quantities correspond to the deformed theory and $\gamma_{ab}$ describes boundary metric. The stress tensor is, by definition, the response of $\CS^{[\mu]}$ to arbitrary variations in the background metric:
	\begin{align}
		\delta \CS^{[\mu]}=\frac{1}{2}\int \d^2x\sqrt{\gamma^{[\mu]}}T^{[\mu]}_{ab}\delta\gamma^{ab}{}^{[\mu]}\,.
	\end{align}
	As described in \cite{Guica:2019nzm}, taking a variation of the defining relation \eqref{TTbar-defn} with respect to the background metric, one obtains after some simple algebra
	\begin{align}
		\del_\mu \gamma_{ab}^{[\mu]}=-2 \hat{T}^{[\mu]}_{ab}~~,~~\del_\mu\hat{T}^{[\mu]}_{ab}=-\hat{T}_{ac}^{[\mu]}\hat{T}^{[\mu]b}_{~~~~c}\,,
	\end{align}
	where $\hat{T}_{ab}=T_{ab}-\gamma_{ab}T$ is the trace reversed stress tensor. These equations are easily solved in terms of the undeformed metric and stress tensors $(\gamma_{ab}^{[0]},\hat{T}^{[0]}_{ab})$ as follows \cite{Guica:2019nzm}
	\begin{align}
		\gamma_{ab}^{[\mu]}&=\gamma_{ab}^{[0]}-2\mu\hat{T}^{[0]}_{ab}+\mu^2\hat{T}^{[0]}_{ac}\hat{T}^{[0]}_{db}\gamma^{[0]\,cd}\,,\notag\\
		\hat{T}^{[\mu]}_{ab}&=T^{[0]}_{ab}-\mu\hat{T}^{[0]}_{ac}\hat{T}^{[0]}_{db}\gamma^{[0]\,cd}\,.\label{MBC-DE-Sol}
	\end{align}
	Note that the above solutions are exact in the deformation parameter, and not perturbative.
	\paragraph{Holography:} In order to obtain the deformed dictionary, we begin with the Fefferman-Graham radial gauge ($g_{\rho a}=0$)
	\begin{align}
		\d s^2=\ell^2\frac{\d \rho^2}{4\rho^2}+g_{ab}\d x^2\d x^b~~,~~g_{ab}=\frac{1}{\rho}g^{(0)}_{ab}+g^{(2)}_{ab}+\rho\,g^{(4)}_{ab}\,.
	\end{align}
	In the case of pure gravity, the undeformed dictionary is given as follows \cite{Skenderis:1999nb}
	\begin{align}
		\gamma_{ab}^{[0]}=g^{(0)}_{ab}~~,~~g^{(2)}_{ab}=8\pi G_N\ell\hat{T}^{[0]}_{ab}~~,~~g^{(4)}_{ab}=\frac{1}{4}g^{(2)}_{ab}g^{(0)\,cd}g^{(0)}_{db}\,.
	\end{align}
	One may now easily find the following non-linear and rather mixed boundary conditions (involving both $g^{(0)}$ and $g^{(2)}$ as compared to only $g^{(0)}$ in the case with Dirichlet boundary conditions) for the $\TTbar$-deformed theory
	\begin{align}
		\gamma_{ab}^{[\mu]}=g_{ab}^{(0)}-\frac{\mu}{4\pi G_N\ell}g^{(2)}_{ab}+\left(\frac{\mu}{4\pi G_N\ell}\right)^2g^{(4)}_{ab}\equiv \rho_c g_{ab}(\rho_c)\,, \label{deformed-dictionary}
	\end{align}
	where $\rho_c=-\frac{\mu}{4\pi G_N\ell}$ and hence $\gamma_{ab}^{[\mu]}$ may be identified as the induced metric on the $\rho=\rho_c$ slice, in accordance with the cut-off prescription in \cite{McGough:2016lol}.
	\subsection{The deformed spacetime: flat background}
Finding the most general solution for the bulk metric compatible with the holographic Ward identities \cite{Skenderis:1999nb} as well as the non-linear mixed boundary conditions in \cref{MBC-DE-Sol,deformed-dictionary} is, in general, a complicated issue. However, for a deformed theory on flat background, it is particularly simple as $g^{(0)}$ and $\gamma^{[\mu]}$ are diffeomorphic to one another \cite{Guica:2019nzm}. In this case, the most general deformed bulk metric may be obtained by applying a two-dimensional coordinate transformation to the generic bulk metric with a flat background ($g^{(0)}=\eta$), namely the Banados metric \cite{Banados:1998gg}
\begin{align}
	\d s^2=\ell^2\frac{\d \rho^2}{4\rho^2}+\frac{\d w\,\d \bar{w}}{\rho}+\CL_\mu(w)\d w^2+\bar\CL_\mu(\bar{w})\d \bar{w}^2+\rho\CL_\mu(w)\bar\CL_\mu(\bar{w})\d w\,\d \bar{w} \,,\label{aux-Banaods}
\end{align}
where $(w,\bar{w})$ denotes the boundary coordinates of the auxiliary flat background discussed above. Therefore, the deformed dictionary \eqref{deformed-dictionary} yields
\begin{align}
	\gamma_{ab}^{[\mu]}\d x^a\d x^b=\left(\d w+\rho_c\bar\CL_\mu(\bar{w})\d \bar{w}\right)\left(\d \bar{w}+\rho_c\CL_\mu(w)\d w\right)\equiv \d Z\,\d \bar Z\,,
\end{align}
where, $(Z,\bar Z)$ are the coordinates which describe the deformed CFT. The two sets of coordinates are related through the following state-dependent coordinate transformations\footnote{The functions $\CL_\mu,\bar\CL_\mu$ are related to the stress tensor expectation values on the auxiliary flat boundary which renders the coordinate transformation \eqref{State-dependent-CT} dynamical. See, for example, \cite{Cardy:2018sdv}.}
\begin{align}
	Z=w+\rho_c\int^{\bar{w}}\bar\CL_\mu(\bar{w})\d \bar{w}~~,~~\bar Z=\bar{w}+\rho_c\int^w\CL_\mu(w)\d w\,.\label{State-dependent-CT}
\end{align}
The Fefferman-Graham expansion of the deformed metric may then be obtained by acting on with the inverse coordinate transformations
\begin{align}
	\begin{pmatrix}
		&\d w\\ &\d \bar{w}
	\end{pmatrix}
	=\frac{1}{1-\rho_c^2\CL_\mu(w)\bar{\CL}_\mu(\bar{w})}\begin{pmatrix}
		1  &-\rho_c\bar{\CL}_\mu(\bar{w})\\
		-\rho_c\CL_\mu(w)  &1
	\end{pmatrix}
	\begin{pmatrix}
		&\d Z\\ &\d \bar Z\,
	\end{pmatrix}\,,\label{inverse-transformations}
\end{align}
on each of the metric coefficients of the auxiliary Banados geometry \eqref{aux-Banaods}.

	\section{Deformed BTZ black hole}\label{sec-3}
	%-------------------------------------------------------------
	%
	A typical example of the deformed spacetime is given by the non-rotating BTZ black hole dual to a $\TTbar$-deformed CFT$_2$ at finite temperature, for which $\CL_\mu=\bar{\CL}_\mu=\textrm{const}$. Utilizing the inverse transformations \eqref{inverse-transformations} and subsequently making the coordinate transformations \cite{Guica:2019nzm,Banerjee:2024wtl}
	\begin{align}
		r=\frac{1+\CL_\mu\rho}{\sqrt{\rho}}~~,~~Z=x+t~~,~~\bar Z=x-t\,,
	\end{align}
	one obtains\footnote{In this article, we set AdS length scale to unity $\ell =1$ and rescale the deformation parameter as $\mu\to\frac{\mu}{8\pi G_N}$.}
	\begin{align}
		\d s^2&=-(r^2-4\CL_\mu)\frac{\d t^2}{(1+2\mu \CL_\mu)^2}+\frac{\d r^2}{r^2-4\CL_\mu}+r^2\frac{\d x^2}{(1-2\mu\CL_\mu)^2}\notag\\
		&x\sim x+2\pi\,.\label{Deformed-BTZ}
	\end{align}
	The event horizon is located at $r=2\sqrt{\CL_\mu}$ and the inverse temperature is given by
	\begin{align}
		\beta=\frac{\pi}{\sqrt{\CL_\mu}}(1+2\mu\CL_\mu)\,.\label{BTZ-temperature}
	\end{align}
	Note that $\beta$ has to be identified with the temperature of the deformed CFT$_2$ and hence the parameter $\CL_\mu$ should be understood as a function of $\mu$ according to the above expression.
	\paragraph{Kruskal extension:}
	Under the coordinate transformations
	\begin{align}
		r=2\sqrt{\CL_\mu}\frac{1-u v}{1+u v}~~,~~t=\frac{1+2\mu\CL_\mu}{4\sqrt{\CL_\mu}}\log\left(-\frac{v}{u}\right)\,,
	\end{align}
	the deformed BTZ black hole metric may be rewritten in the Kruskal form
	\begin{align}
		\d s^2=-\frac{4\d u\d v}{(1+u v)^2}+4\CL_\mu\left(\frac{1-u v}{1+u v}\right)^2\frac{\d x^2}{(1-2\mu\CL_\mu)^2}\,.\label{Kruskal-metric}
	\end{align}
	Note that the Kruskal coordinates $(u,v,x)$ smoothly cover the maximally extended black hole spacetime, dual to the thermofield double state in the deformed CFT$_2$ \cite{Maldacena:2001kr}. The boundaries are at $u v=-1$, while the event horizons and the future and past singularities are given, respectively by $uv=0$ and $uv=1$. The inverse coordinate transformations may also be obtained for the left and right geometries, respectively, as follows
	\begin{align}
		&\textrm{(left)}~~u=e^{-\frac{2\sqrt{\CL_\mu}}{1+2\mu \CL_\mu}t}\sqrt{\frac{r-2\sqrt{\CL_\mu}}{r+2\sqrt{\CL_\mu}}}~~,~~v=-e^{\frac{2\sqrt{\CL_\mu}}{1+2\mu \CL_\mu}t}\sqrt{\frac{r-2\sqrt{\CL_\mu}}{r+2\sqrt{\CL_\mu}}}\,,\notag\\
		&\textrm{(right)}~~u=-e^{-\frac{2\sqrt{\CL_\mu}}{1+2\mu \CL_\mu}t}\sqrt{\frac{r-2\sqrt{\CL_\mu}}{r+2\sqrt{\CL_\mu}}}~~,~~v=e^{\frac{2\sqrt{\CL_\mu}}{1+2\mu \CL_\mu}t}\sqrt{\frac{r-2\sqrt{\CL_\mu}}{r+2\sqrt{\CL_\mu}}}\,.\label{left-right-uv}
	\end{align}
	\paragraph{Embedding in \(\mathbb{R}^{2,2}\):}
	The deformed geometry is still asymptotically AdS$_3$ (maximally symmetric) and therefore can be embedded in \(\mathbb{R}^{2,2}\) with a quadratic constraint as follows
	\begin{align}
		ds^2 =& \eta_{AB}\d X^A\d X^B~~,~~X^2\equiv \eta_{AB}X^AX^B=-1\,,
	\end{align}
	where the standard metric on $\BR^{2,2}$ is given by $\eta=\textrm{diag}(-1,-1,1,1)$.
	For example, an embedding ansatz is given by \cite{Shenker:2013pqa}:
	\begin{align}
		X_1&=\frac{v+u}{1+u v}=\sqrt{\frac{r^2}{4\CL_\mu}-1}\sinh\left(\frac{2\sqrt{\CL_\mu}t}{1+2\mu \CL_\mu}\right)\,,\notag\\
		X_2&=\frac{1-u v}{1+u v}\cosh\left(\frac{2\sqrt{\CL_\mu}x}{1-2\mu \CL_\mu}\right)=\frac{r}{2\sqrt{\CL_\mu}}\cosh\left(\frac{2\sqrt{\CL_\mu}x}{1-2\mu \CL_\mu}\right)\,,\notag\\
		X_3&=\frac{v-u}{1+u v}=\sqrt{\frac{r^2}{4\CL_\mu}-1}\cosh\left(\frac{2\sqrt{\CL_\mu}t}{1+2\mu \CL_\mu}\right)\,,\notag\\
		X_4&=\frac{1-u v}{1+u v}\sinh\left(\frac{2\sqrt{\CL_\mu}x}{1-2\mu \CL_\mu}\right)=\frac{r}{2\sqrt{\CL_\mu}}\sinh\left(\frac{2\sqrt{\CL_\mu}x}{1-2\mu \CL_\mu}\right)\,.\label{embedding}
	\end{align}
	The geodesic length between two arbitrary bulk points $X^A\equiv(u,v,x)$ and $X^A\equiv(u^\prime,v^\prime,x^\prime)$ may be computed in the embedding coordinates, as
	\begin{align}
		d(X,X^\prime)&=\textrm{arccosh}(-X\cdot X^\prime)\notag\\
		&=\textrm{arccosh}\left[\frac{2(u v^\prime+v u^\prime)+(1-u v)(1-u^\prime v^\prime)\cosh\left[\frac{2\sqrt{\CL_\mu}(x-x^\prime)}{1-2\mu\CL_\mu}\right]}{(1+u v)(1+u^\prime v^\prime)}\right]\,.\label{Geodesic-distance}
	\end{align}
	where the dot product is taken with respect to the metric $\eta_{AB}$.
	%-------------------------------------------------------------
	%
	\subsection{Spherically symmetric shockwave}
	In this subsection, following \cite{Shenker:2013pqa}, we perturb the dual thermofield double state (TFD) by adding a few quantas at the left boundary and analyze the effects of the gravitational shockwave in the deformed bulk geometry. Our goal is to study the out-of-time-ordered correlator (OTOC) in the TFD state:
	\begin{align}
		\left<W(t_2,x_2)V(t_1,x_1)W(t_2,x_2)V(t_1,x_1)\right>_\beta\,,
	\end{align}
	where $t_2-t_1\gg\beta$. In the Kruskal coordinates, the OTOC may be interpreted as a gravitational scattering problem between the $W$ and $V$ particles, whose dual operators are inserted in the boundary state \cite{Shenker:2013pqa,Roberts:2014ifa}.  
	
\begin{figure}[ht]
		\centering
		\includegraphics[width=0.65\textwidth]{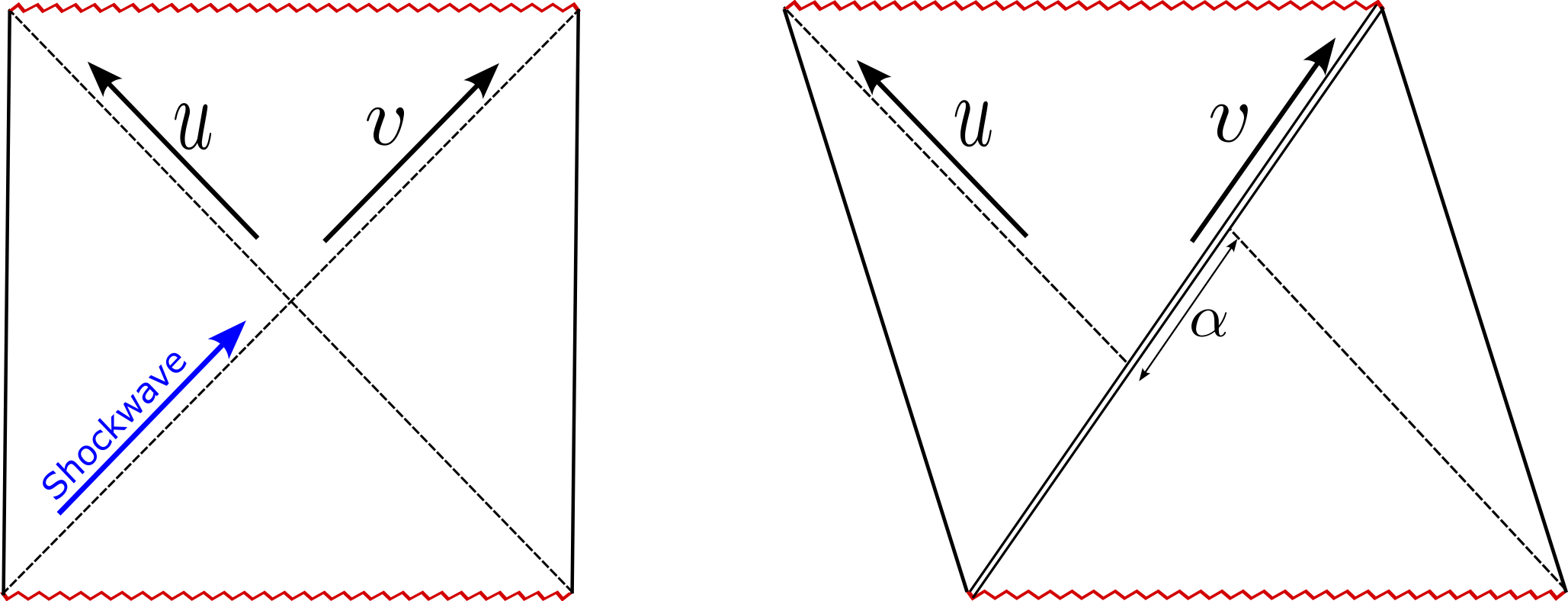}
		\caption{(Left panel) Insertion of a few particles from the left boundary leads to a gravitational shockwave in the bulk geometry. (Right panel) The Penrose diagram after the shockwave insertion, depicted by the double line. The dashed $v=0$ and $\bar v=0$ horizons miss by an amount $\alpha$.}
		\label{fig:shock-insertion}
	\end{figure}

	Now, we mildly perturb the deformed black hole spacetime by adding a few particles at the left-boundary, and let them fall into the black hole as depicted in \cref{fig:shock-insertion}. As customary, we shall designate these particles by $W$. Although the particles have a very small energy which does not seem to perturb the background geometry significantly, one should note the following:
	\paragraph{Fact:} If a particle released from the boundary $r\to\infty$ at an early time $t_W$ is moving along a null trajectory with proper energy $E$, then the energy $E_r$ measured on the time slice $t=0$ is given by
	\begin{align}
		E_r=\frac{E}{\sqrt{g_{00}\big|_{t=0}}}\,.
	\end{align}
	For the deformed BTZ black hole \eqref{Deformed-BTZ}, the null trajectories may be found by defining the Tortoise coordinate\footnote{In terms of the tortoise coordinate, the Kruskal coordinates may be rewritten in the familiar form,
	\begin{align}
		u= -e^{-\frac{2\sqrt{\CL_\mu}}{1+2\mu\CL_\mu}(t-r_\star)}~~,~~v=e^{\frac{2\sqrt{\CL_\mu}}{1+2\mu\CL_\mu}(t+r_\star)}\,.
		\end{align}} as  follows
	\begin{align}
		\d r_\star=(1+2\mu\CL_\mu)\frac{\d r}{r^2-4\CL_\mu}\implies r_\star=\frac{(1+2\mu\CL_\mu)}{4\sqrt{\CL_\mu}}\log\left(\frac{r-2\sqrt{\CL_\mu}}{r+2\sqrt{\CL_\mu}}\right)\,.\label{Tortoise}
	\end{align}
	Therefore, the null trajectories are given by
	\begin{align}
		&-(r^2-4\CL_\mu)\frac{\d t^2}{(1+2\mu\CL_\mu)^2}+\frac{\d r^2}{r^2-4\CL_\mu}=0\,,\notag\\
		\implies & \int_{t_W}^t\d t=\int_{0}^{r_\star}\d r_\star \implies t=t_W+r_\star \,.
	\end{align}
	Hence, the blue-shifted energy at the time slice $t=0$ is given by
	\begin{align}
		E_r=\frac{1+2\mu \CL_\mu}{\sqrt{r^2-4\CL_\mu}\big|_{t=0}}E\sim \frac{E(1+2\mu \CL_\mu)}{2\sqrt{\CL_\mu}}e^{\frac{2\sqrt{\CL_\mu}}{1+2\mu\CL_\mu}t_W}\,.
	\end{align}
	From the above expression it is clear that, if $t_W$ is sufficiently large, we must include the effects of backreaction of this energy. Similar to the undeformed case described in \cite{Shenker:2013pqa}, we construct the geometry by gluing a deformed BTZ metric of mass $M$ to  that of mass $M+E$ across the null surface
		\begin{align}
			u_W=e^{-\frac{2\sqrt{\CL_\mu}}{1+2\mu\CL_\mu}t_W}\,,
		\end{align}
		where $E\ll M$ is the asymptotic energy of the perturbation. We will use coordinates $(u,v)$ to the right of the shell and $(\hat u,\hat v)$ to the left as depicted in \cref{fig:shock-insertion}.  It is straightforward to verify that the horizon radii scales as 
		\begin{align}
			\hat \CL_\mu=\frac{E+M}{M}\CL_\mu\,.
		\end{align}
	Similar to the undeformed case \cite{Shenker:2013pqa}, we will use the matching conditions
	\begin{enumerate}
		\item The time coordinate $t$ flows continuously across the shell.
		\item The radius of the $S^1$ be continuous across the shell.
	\end{enumerate}
	The first condition determines the location of the shell in terms of the hatted coordinates as
	\begin{align}
		\hat u_W=e^{-\frac{2\sqrt{\hat\CL_\mu}}{1+2\mu\hat\CL_\mu}t_W}\,,
	\end{align}
	whereas the second condition implies the following relation
	\begin{align}
		\frac{\sqrt{\hat\CL_\mu}}{1-2\mu\hat\CL_\mu}\frac{1-\hat u_W\hat v}{1+\hat u_W\hat v}=\frac{\sqrt{\CL_\mu}}{1-2\mu\CL_\mu}\frac{1-u_W v}{1+u_W v}\,.
	\end{align}
	For small $\frac{E}{M}$, the above relation may be solved straightforwardly to obtain
	\begin{align}\label{alphavalue}
		\hat v=v+\alpha~~,~~\alpha=\frac{E}{4M}\frac{1+2\mu\CL_\mu}{1-2\mu\CL_\mu}e^{\frac{2\sqrt{\CL_\mu}}{1+2\mu\CL_\mu}t_W}\,.
	\end{align}
	This matching condition is exact in the limit $\frac{E}{M}\to 0$ and $t_W\to \infty$ with $\alpha$ fixed. Furthermore, in this limit, we have $\hat\CL_\mu=\CL_\mu$ and hence the bulk metric may be written as
	\begin{align}
		\d s^2=\frac{-4\d u\d v}{\left[1+u(v+\alpha\Theta(u))\right]^2}+4\CL_\mu\left[\frac{1-u(v+\alpha\Theta(u))}{1+u(v+\alpha\Theta(u))}\right]^2\frac{\d x^2}{(1-2\mu\CL_\mu)^2}\,.
	\end{align}
	In terms of the discontinuous coordinates $\bar u=u\,,\,\bar v=v+\alpha\Theta(u)$, the metric takes a more standard shockwave form
	\begin{align}
		\d s^2=\frac{-4\d \bar u\d \bar v+4\alpha\delta(\bar u)\d \bar u^2}{(1+\bar u \bar v)^2}+4\CL_\mu\left(\frac{1-\bar u \bar v}{1+\bar u \bar v}\right)^2\frac{\d x^2}{(1-2\mu\CL_\mu)^2}\,.
	\end{align}
%	\begin{figure}[ht]
%		\centering
%		\includegraphics[width=0.75\textwidth]{Penrose.png}
%		\caption{The Kruskal and Penrose diagrams of the geometry with a shockwave from the left asymptotic boundary, depicted by the double line. The dashed $v=0$ and $\bar v=0$ horizons miss by an amount $\alpha$.}
%		\label{fig:Shockwave-Kruskal-Penrose}
%	\end{figure}
	It is straightforward to check that Einstein equations imply a stress tensor 
	\begin{align}
		T_{uu}=\frac{\alpha}{4\pi G_N}\delta(u)\,,
	\end{align}
	corresponding to a shell of null particles symmetrically distributed on the horizon.
	%-------------------------------------------------------
	\subsubsection{Geodesics and Shockwave Effects: OTOC}
	Consider a geodesic connecting a point at Killing time $t_L$ on the left boundary  with a point at Killing time $t_R$ on the right boundary. We will take both points to be located at the same value of $x=x_0$. As depicted in the figure below, such a geodesic is made up of two geodesic segments on the left and right of the shockwave, respectively, which are joined smoothly at the horizon $u=0$.
	\begin{figure}[ht]
		\centering
		\includegraphics[width=0.5\textwidth]{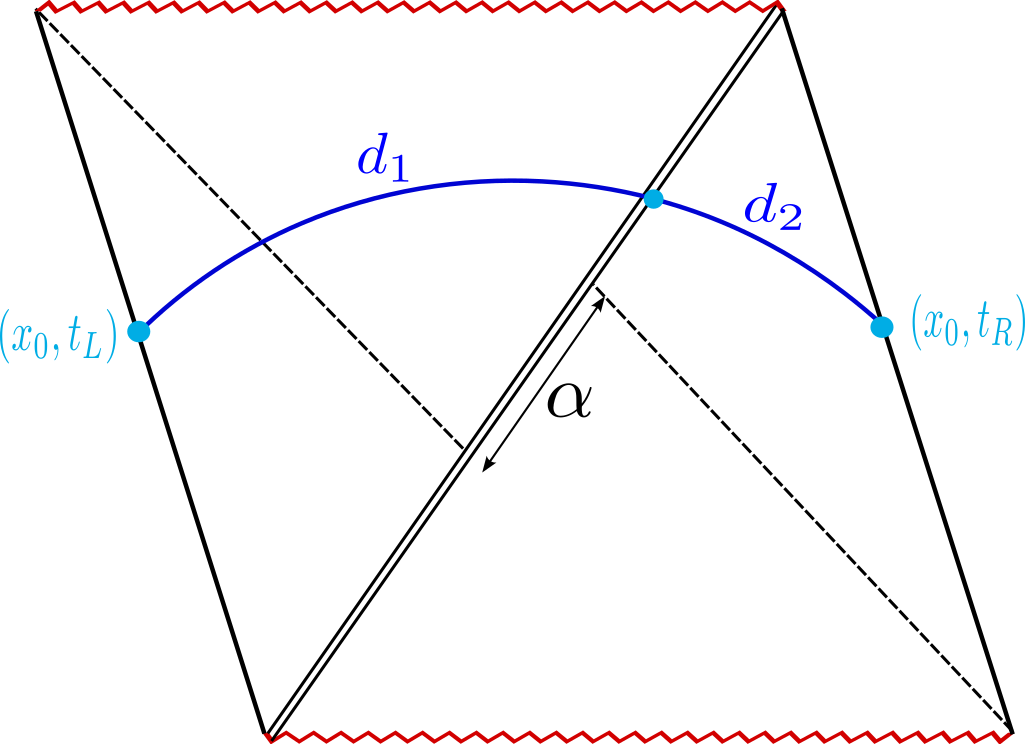}
		\caption{Computing the geodesic distance between two points $(x_0,t_L)$ and $(x_0,t_R)$ on the left and right asymptotic boundaries in the shockwave geometry.}
		\label{fig:geodesic}
	\end{figure}
	In \cref{fig:geodesic}, $d_1$ corresponds to the length
	between the boundary point $(u^L_\infty,v^L_\infty,x_0)$ (the subscript $\infty$ is used to denote the boundary point) and an arbitrary point $(0,v+\alpha,x)$ on the horizon. Similarly, $d_2$ denotes the geodesic length between $(0,v,x)$ and $(u^R_\infty,v^R_\infty,x_0)$ on the right Kruskal geometry. Using the geodesic distance formula \eqref{Geodesic-distance} in embedding coordinates, we may now obtain
	\begin{align}
		d_1=\textrm{arccosh}\left[\frac{r}{2\sqrt{\CL_\mu}}\cosh\frac{2\sqrt{\CL_\mu}(x-x_0)}{1-2\mu\CL_\mu}+\frac{\sqrt{r^2-4\CL_\mu}}{2\sqrt{\CL_\mu}}(v+\alpha)e^{-\frac{2\sqrt{\CL_\mu}t_L}{1+2\mu\CL_\mu}}\right]\,,
	\end{align}
	where, we have used the fact that,on the left geometry
	\begin{align*}
		u_L=\frac{\sqrt{r^2-4\CL_\mu}}{r+2\sqrt{\CL_\mu}}e^{-\frac{2\sqrt{\CL_\mu}t_L}{1+2\mu\CL_\mu}}~~,~~v_L=-\frac{\sqrt{r^2-4\CL_\mu}}{r+2\sqrt{\CL_\mu}}e^{\frac{2\sqrt{\CL_\mu}t_L}{1+2\mu\CL_\mu}}\,.
	\end{align*}
	In a similar manner,
	\begin{align}
		d_2=\textrm{arccosh}\left[\frac{r}{2\sqrt{\CL_\mu}}\cosh\frac{2\sqrt{\CL_\mu}(x-x_0)}{1-2\mu\CL_\mu}-\frac{\sqrt{r^2-4\CL_\mu}}{2\sqrt{\CL_\mu}}v\, e^{-\frac{2\sqrt{\CL_\mu}t_R}{1+2\mu\CL_\mu}}\right]\,.
	\end{align}
	For large $r=r_\infty$, from \cref{left-right-uv} we may obtain the total length of the two segments as
	\begin{align}
		d=\log\left[\frac{r_\infty^2}{\CL_\mu}\left(\cosh\frac{2\sqrt{\CL_\mu}(x-x_0)}{1-2\mu\CL_\mu}+(v+\alpha)e^{-\frac{2\sqrt{\CL_\mu}t_L}{1+2\mu\CL_\mu}}\right)\left(\cosh\frac{2\sqrt{\CL_\mu}(x-x_0)}{1-2\mu\CL_\mu}-v\,e^{-\frac{2\sqrt{\CL_\mu}t_R}{1+2\mu\CL_\mu}}\right)\right]\,.
	\end{align}
	Now extremizing the total length over the arbitrary parameters $(x,v)$, we obtain
	\begin{align}
		x=x_0~~,~~v=\frac{1}{2}\left(-e^{\frac{2\sqrt{\CL_\mu}t_L}{1+2\mu\CL_\mu}}+e^{\frac{2\sqrt{\CL_\mu}t_R}{1+2\mu\CL_\mu}}-\alpha\right)\,,
	\end{align}
	leading to the minimal geodesic distance
	\begin{align}
		d_\textrm{min}&=2\log\left[\frac{r_\infty}{\sqrt{\CL_\mu}}\left(\cosh\frac{\sqrt{\CL_\mu}(t_L-t_R)}{1+2\mu\CL_\mu}+\frac{\alpha}{2}\,e^{-\frac{\sqrt{\CL_\mu}}{1+2\mu\CL_\mu}(t_L+t_R)}\right)\right]\notag\\
		&=2\log\left[\frac{\beta-\sqrt{\beta^2-8\pi^2\mu}}{4\pi\mu\,\epsilon_c}\left(\cosh\frac{\pi(t_L-t_R)}{\beta}+\frac{\alpha}{2}\,e^{-\frac{\pi(t_L+t_R)}{\beta}}\right)\right]\,,\label{geodesic-distance-LR}
	\end{align}
	where we have used the cut-off $\epsilon_c=1/r_\infty$ for large $r$.
	The first term in the parenthesis is the usual Hartman-Maldacena contribution \cite{Hartman:2013qma}, whereas the second term, proportional to $\alpha$ corresponds to the effect of the shockwave. Note that for large $t_L,t_R$ the effect of the shockwave on the geodesic distance becomes negligible as in the undeformed case.
	
	\subsubsection*{OTOC}
	In the geodesic approximation, the OTOC may be obtained as the two point function of the (quasi-primary) fields $V_L$ and $V_R$ in the left and right CFTs respectively, computed in the perturbed geometry where a $W$ operator creates a few particles at the far past $t_W$. Assuming that the saddle point approximation still holds for the gravitational path integral, we may obtain this correlation function in the geodesic approximation in terms of a particle of mass $m$ following a geodesic trajectory in the bulk \cite{Shenker:2013pqa}:
	\begin{align}
		\left<W(t_W)V_L(0)V_R(0)W(t_W)\right>\equiv \left<W(t_W)|V_LV_R|W(t_W)\right>\approx e^{-m^{}_W d}\,.
	\end{align}
	For $t_L=t_R=0$, we have the following estimate for the  OTOC:
	\begin{align}
		\frac{\left<V_LV_R\right>_W}{\left<W W\right>\left<V_LV_R\right>}\sim&\left[\frac{1}{1+\frac{E}{8M}\frac{1+2\mu\CL_\mu}{1-2\mu\CL_\mu}e^{\frac{2\sqrt{\CL_\mu}}{1+2\mu\CL_\mu} t_W}}\right]^{2m^{}_W}\,,
		%\\
		%\frac{\left<V_LV_R\right>_W}{\left<W W\right>\left<V_LV_R\right>}\sim&\left[1-2m \frac{E}{8 M}\frac{1+2\mu\CL_\mu}{1-2\mu\CL_\mu}e^{\frac{2\sqrt{\CL_\mu}}{1+2\mu\CL_\mu} t_W}\right]\,,
	\end{align}
	which enables us to identify the Lyapunov exponent as (cf. \cref{BTZ-temperature})
	\begin{align}
		\lambda_L=\frac{2\sqrt{\CL_\mu}}{1+2\mu\CL_\mu}=\frac{2\pi}{\beta}\,.
	\end{align}
	Therefore, we may conclude that the MSS bound \cite{Maldacena:2015waa} is saturated even for the deformed CFT$_2$.

	\subsection{Localized Shock}
	For a more realistic scenario, in this section, we consider a localized shockwave introduced in \cite{Dray:1984ha, Dray:1985yt} \footnote{For further studies in similar geometries, see also \cite{Hotta:1992qy, Sfetsos:1994xa, Cai:1999dz, Cornalba:2006xk}.}. In particular, the authors of \cite{Dray:1984ha} considered spherical shockwave generated by a massless particle, moving along the horizon of a Schwarzschild black hole. This shockwave is localized on the null hypersurface at the horizon ($u=0$), causing a coordinate shift in the null coordinate $v$.
	The metric of a generic two-sided static black hole in null coordinates is given by:
	\begin{align}
		ds^2 = 2 A(u,v) \d u\,\d v + B(u,v)\,\d x^2.\label{generic}
	\end{align}
	In the present scenario, from \cref{Kruskal-metric}, we may identify
	\begin{align}
		A(u,v)=\frac{-2}{(1+u v)^2}~~,~~B(u,v)=\frac{4\CL_\mu}{(1-2\mu\CL_\mu)^2}\left(\frac{1-u v}{1+u v}\right)^2\,.\label{A-B-defn}
	\end{align}
	The (unperturbed) background obeys the Einstein equations
	\begin{align}
		R_{\mu\nu}-\frac{1}{2}g_{\mu\nu}R+\Lambda g_{\mu\nu}=\kappa_G T_{\mu\nu}~~,~~\kappa_G=8\pi G_N\,,
	\end{align}
	where the matter stress tensor may be obtained as follows
	\begin{align}
		T_{\mu\nu}^\textrm{matter}\d x^\mu \d x^\nu=T_{uu}\d u^2+2T_{uv}\d u\,\d v+T_{vv}\d v^2+T_{xx}\d x^2\,.\label{unbarred-matter}
	\end{align}
	The components of the matter stress tensor may be obtained from the Einstein equations, in terms of the functions $A\,,\,B$ as follows
	\begin{align}
		&\kappa_G T_{uu}=\frac{A\del_u B+2B(\del_u A\del_u B-A\del_u^2B)}{4AB^2}\,,\notag\\
		&\kappa_G T_{uv}=\frac{\del_u A\del_v A-A\del_u\del_v A}{A^2}+\frac{\del_u B\del_v B-2B\del_u\del_v B}{4B^2}+2A\,,\notag\\
		&\kappa_G T_{vv}=\frac{A\del_v B+2B(\del_v A\del_v B-A\del_v^2B)}{4AB^2}\,,\notag\\
		&\kappa_G T_{xx}=\frac{\del_u B\del_v B-2B\del_u\del_v B}{2AB}+2B\,.
	\end{align}
	In the following, we will investigate the effects of inserting a shockwave in the background \eqref{generic} and subsequently compute the OTOC in geodesic approximation.
	\subsubsection{Backreaction due to shockwave}
	We consider the shockwave with stress-energy tensor
	\begin{align}
		T_{uu}^{\textrm{shock}}=E_0\,e^{\frac{2\pi t}{\beta}}\delta(u)\delta(x)\,,
	\end{align}
	localized on the horizon $u=0$ at $x=0$. Here, $E_0$ is a constant related to the asymptotic energy of the shockwave and $e^{\frac{2\pi t}{\beta}}$ represents the blue-shift factor.
	\begin{figure}[ht]
		\centering
		\includegraphics[width=0.65\textwidth]{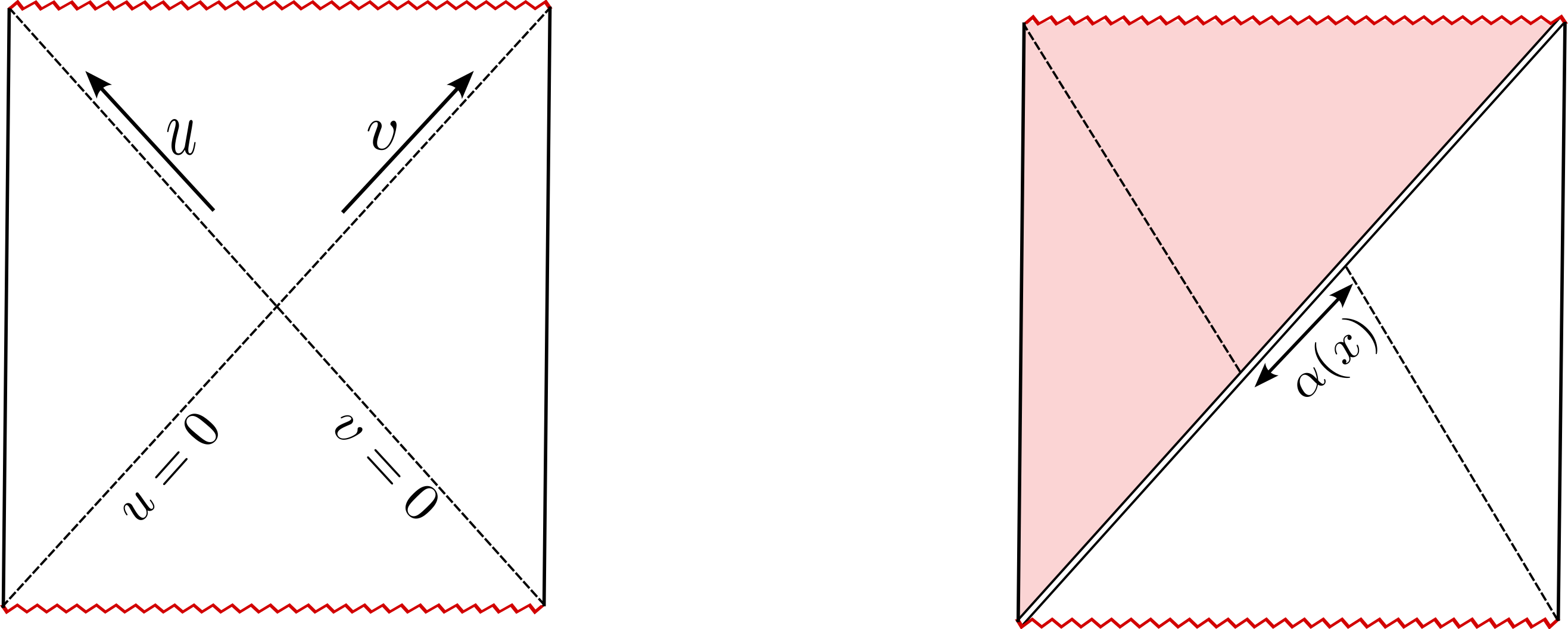}
		\caption{A localized shock.}
		\label{fig:localized-shock}
	\end{figure}
	Upon introduction of the shock, the metric in the red region in \cref{fig:localized-shock} will change. The shockwave geometry is characterized by the shift \cite{Dray:1984ha,Dray:1985yt,Sfetsos:1994xa}
	\begin{align}
		u\to u~~&,~~v\to v+\Theta(u)\alpha(t,x)\,,\notag\\
		\d u\to \d u~~&,~~\d v\to \d v+\alpha(t,x)\delta(u)\d u\,.
	\end{align}
	As earlier, we may redefine the Kruskal coordinates as \cite{Shenker:2013pqa,Roberts:2014isa}
	\begin{align}
		\bar u=u~,~\bar v=v+\Theta(u)\alpha(t,x)~,~\bar x=x\,.
	\end{align}
	In terms of the new coordinates, the shockwave metric looks like
	\begin{align}
		\d s^2=2A(\bar u,\bar v)\d\bar u\,\d\bar v+B(\bar u,\bar v)\d\bar x^2-2A(\bar u,\bar v)\alpha(\bar x)\delta(\bar u)\d\bar u^2\,.
	\end{align}
	The backreacted Einstein equations take the form
	\begin{align}
		R_{\mu\nu}-\frac{1}{2}g_{\mu\nu}R+\Lambda g_{\mu\nu}=\kappa\left(T_{\mu\nu}^\textrm{matter}+T_{\mu\nu}^\textrm{shock}\right)\,,
	\end{align}
	where the matter part takes the following form in the barred coordinates (cf. \cref{unbarred-matter})
	\begin{align}
		T_{\mu\nu}^\textrm{matter}\d x^\mu\d x^\nu=&\left(T_{\bar u\bar u}+\alpha(t,x)^2\delta(u)^2T_{\bar v\bar v}-2\alpha(t,x)\delta(u)T_{\bar u\bar v}\right)\d\bar u^2\notag\\&+2\left(T_{\bar u\bar v}-2\alpha(t,x)\delta(u)T_{\bar v\bar v}\right)\d\bar u\d\bar v+T_{\bar v\bar v}\d\bar v^2+T_{\bar x\bar x}\d\bar x^2\,.
	\end{align}
	To keep track of the order of perturbation we multiply the source and the effect of geometric shift with a constant book keeping parameter $\sigma$:
	\begin{align*}
		T_{\mu\nu}^\textrm{shock}\to\sigma T_{\mu\nu}^\textrm{shock}~~,~~\alpha\to \sigma\alpha\,.
	\end{align*}
	Then the $\bar v\bar v$ component of the Einstein equations at the linear order in $\sigma$ gives the following differential equation
	\begin{align}
		&\left(\del_x^2-\frac{\del_u\del_vB}{2A}\right)\alpha(x)=8\pi G_N \frac{B}{A}E_0e^{\frac{2\pi t}{\beta}}\delta(x)\,,\notag\\
		\textrm{or,}\,&\left(\del_x^2-\CM(u,v)^2\right)\alpha(x)=e^{\frac{2\pi}{\beta}(t-t_\star)}\delta(x)\,,
	\end{align}
	where, we have defined
	\begin{align}
		\CM(u,v)=\sqrt{\frac{\del_u\del_vB}{2A}}~~,~~t_\star=\frac{\beta}{2\pi}\log\left(\frac{A}{8\pi G_N E_0 B}\right)\,.
	\end{align}
	Our job is to solve the above equation near the horizon $u=0$, whence, using \cref{A-B-defn}:
	\begin{align}
		\CM^2=\frac{\del_u\del_vB}{2A}\Big|_{u=0}=\frac{4\pi^2}{\beta^2-8\pi^2\mu}~~,~~t_\star=\frac{\beta}{2\pi}\log\left(\frac{\beta ^2-8 \pi ^2 \mu }{16 \pi ^3G_N E_0}\right)\,.
	\end{align}
	The homogeneous equation has the following solution:
	\begin{align}
		\alpha(x)=\begin{cases}
			c_1 e^{\CM x}+c_2e^{-\CM x}~~&\textrm{for}\,\, x>0\,,\notag\\
			c_3 e^{\CM x}+c_4e^{-\CM x}~~&\textrm{for}\,\, x<0\,.
		\end{cases}
	\end{align}
	The delta function contributes to the first order discontinuity:
	\begin{align}
		\alpha^\prime(\epsilon)-\alpha^\prime(-\epsilon)=e^{\frac{2\pi}{\beta}(t-t_\star)}\,,
	\end{align}
	where $\epsilon\to 0$ is a small parameter. These conditions lead to 
	\begin{align}
		c_1-c_3=c_4-c_2=\frac{1}{2\CM}e^{\frac{2\pi}{\beta}(t-t_\star)}\,.
	\end{align}
	We may choose $c_2=c_3=0$ \cite{Shenker:2014cwa}, to obtain the final profile of the shockwave as
	\begin{align}
		\alpha(x)=\frac{1}{2\CM}e^{\frac{2\pi}{\beta}\left(t-t_\star-\frac{\beta\CM}{2\pi}x\right)}\,.
	\end{align}
	\subsubsection{Violation of the Mezei-Stanford bound}
\begin{figure}[ht]
		\centering
		\includegraphics[width=0.35\textheight]{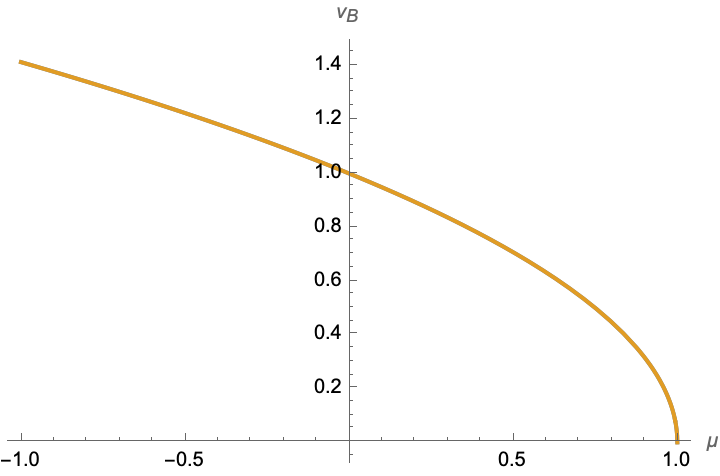}
		\caption{Variation of the butterfly velocity with respect to the deformation parameter $\mu$ in the deformed BTZ black hole geometry. For $\mu<0$, the butterfly velocity violated the bound derived in \cite{Mezei:2016wfz}.} 
		\label{fig:vB-BTZ}
	\end{figure}

	A straightforward analysis of a geodesic connecting two points on the left and right asymptotic boundaries of the shockwave geometry will lead to the OTOC:
	\begin{align}
		\frac{\left<V_LV_R\right>_W}{\left<W W\right>\left<V_LV_R\right>}&\sim\left[\frac{1}{1+e^{\frac{2\pi}{\beta}\left(t-t_\star-\frac{\beta\CM}{2\pi}x\right)}}\right]^{2m^{}_W}\,,
		%\\
		%\frac{\left<V_LV_R\right>_W}{\left<W W\right>\left<V_LV_R\right>}&\sim\left[1-2m e^{\frac{2\pi}{\beta}\left(t-t_\star-\frac{\beta\CM}{2\pi}x\right)}\right]\,.
	\end{align}
	Therefore, the Lyapunov exponent and the butterfly velocity may be read off as follows
	\begin{align}
		\lambda_L=\frac{2\pi}{\beta}~~,~~v_B=\frac{2\pi}{\beta\CM}=\sqrt{1-\frac{8\pi^2\mu}{\beta^2}}\,.\label{vB-BTZ}
	\end{align}
	It is easy to see that although for $\mu>0$, the Mezei-Stanford bound is satisfied, we see a clear violation of the bound for negative values of the deformation parameter (cf. \cref{fig:vB-BTZ}) which allows for superluminal propagation of chaos in $\TTbar$ deformed theories. This result, although seemingly violating causality, is not very surprising. As discussed in \cite{McGough:2016lol,Cardy:2018sdv},  the signal propagation speed in $\TTbar$ deformed theories can become superluminal for\footnote{Note the difference in the convention regarding the signature of $\mu$ in \cite{McGough:2016lol,Cardy:2018sdv} as compared to ours.} $\mu<0$, due to the repulsive nature of the particle interactions. This may also be understood as the $\TTbar$ deformation effectively placing the CFT on a curved manifold (cf. \cref{AppA2,AppB}), whose causal structure provides the notion of a rescaled butterfly velocity \cite{McGough:2016lol}. In fact, the bound $v_B\leq 1$ obtained in \cite{Qi:2017ttv} for asymptotically AdS spacetimes obeying the null energy condition, has been shown to be violated for non-local boundary theories in \cite{Fischler:2018kwt}. Further supporting evidences of superluminal butterfly velocity may be found in \cite{Das:2021qsd,Das:2022jrr} where the authors considered the OTOC in driven and quenched CFTs.
	
	Interestingly, near the critical value of $\mu$ originating from the reality of the ground state energy, namely the Hagedorn bound \cite{Cavaglia:2016oda,Dubovsky:2012wk,McGough:2016lol}
	\begin{align}
		\mu< \mu_c=\frac{\beta^2}{8\pi}\,,
	\end{align}
	the butterfly velocity approaches zero, indicating hindered spread of chaos in $\TTbar$-deformed theories. This phenomena, perhaps, also has its origin in the non-local nature of such theories \cite{McGough:2016lol}.
	%In certain regimes, the butterfly velocity \(v_B\) may exceed bounds predicted by the Mezei-Stanford analysis:
	
	\subsection{Pole-skipping}
	In this section, we wish to analyze the retarded energy density Green's function $G^R_{T^{00}T^{00}}(\omega,k)$ holographically, utilizing the techniques developed in \cite{Blake:2017ris,Blake:2018leo}. This correlation function may be analyzed by perturbing the gravitational background and imposing ingoing boundary conditions near the horizon. It is customary to utilize the ingoing Eddington-Finkelstein coordinates
	\begin{align}
		\mathtt{v}=t+r_\star~~,~~\d r_\star=(1+2\mu\CL_\mu)\frac{\d r}{r^2-4\CL_\mu}\,.
	\end{align}
	From \cref{Tortoise}, we may therefore obtain
	\begin{align}
		\mathtt{v}=t+\frac{1+2\mu \CL_\mu}{4\sqrt{\CL_\mu}}\log\left(\frac{r+2\sqrt{\CL_\mu}}{r-2\sqrt{\CL_\mu}}\right)\,,
	\end{align}
	and subsequently the metric \eqref{Deformed-BTZ} may be written in the ingoing EF gauge as follows
	\begin{align}
		\d s^2=-\left(r^2-4 L\right) \frac{\d \mathtt v^2}{(1+2 \mu  \CL_\mu)^2}-\frac{2 \d r\, \d\mathtt v}{1+2 \mu  \CL_\mu}+r^2 \frac{\d x^2}{(1-2 \mu  \CL_\mu)^2}\,.\label{dBTZ-in-EF}
	\end{align}
	The action for $3d$ Einstein gravity is
	\begin{align}
		S_\textrm{EH}=\frac{1}{16\pi G_N}\int \d^3x\sqrt{-g}(R-2\Lambda)\,,
	\end{align}
	and the equations of motion following from the above action are given by (recall that $\Lambda$ is related to the AdS radius as $\Lambda=-\frac{1}{\ell^2}$ and we set $\ell=1$)
	\begin{align}
		E_{\mu\nu}=R_{\mu\nu}-\frac{1}{2}g_{\mu\nu}R-g_{\mu\nu}=0\,.\label{EOM}
	\end{align}
	It is straightforward to verify that \eqref{dBTZ-in-EF} satisfies \eqref{EOM}. 
	
	To obtain the energy density Green's function in the momentum space, we now perturb the background metric by putting perturbations on the longitudinal modes or the sound modes. In particular, we choose the ansatz \cite{Blake:2018leo}
	\begin{align}
		g_{\mu\nu}\to g_{\mu\nu}+h_{\mu\nu}=g_{\mu\nu}+\delta g_{\mu\nu}(r)\,e^{-i\left(\omega \mathtt v-k x\right)}\,,\label{logitudinal-perturbations}
	\end{align}
	where $g_{\mu\nu}$ satisfies the vacuum Einstein's equations. Plugging the above ansatz into \eqref{EOM}, we may obtain the field equations for $h_{\mu\nu}$ as follows \cite{Liu:2020yaf}
	\begin{align}
		E_{\mu\nu}=\nabla^\rho\nabla_{(\mu} h_{\nu)\rho}-\frac{1}{2}\nabla^\rho\nabla_\rho h_{\mu\nu}-\frac{1}{2}\nabla_\mu\nabla_\nu h&-\frac{1}{2}g_{\mu\nu}\nabla^\rho\nabla^\sigma h_{\rho\sigma}\notag\\
		&+\frac{1}{2}g_{\mu\nu}\nabla^\rho\nabla_\rho h+2h_{\mu\nu}-2h g_{\mu\nu}=0\,,
	\end{align}
	where $h=g^{\mu\nu}h_{\mu\nu}$. The above equation can also be obtained from the second order expansion of the Einstein-Hilbert action around the background \eqref{dBTZ-in-EF} \cite{Liu:2020yaf}:
	\begin{align}
		S=\frac{1}{16\pi G_N}\int \d^3 x\sqrt{-g}\Bigg[2h_{\mu\nu}h^{\mu\nu}-h^2&-\frac{1}{2}\left(\nabla_\mu h\right)\left(\nabla^\mu h\right)+\frac{1}{2}\left(\nabla_\rho h_{\mu\nu}\right)\left(\nabla^\rho h^{\mu\nu}\right)\notag\\
		&+\left(\nabla_\mu h\right)\left(\nabla_\nu h^{\mu\nu}\right)-\left(\nabla_\nu h_{\mu\rho}\right)\left(\nabla^\rho h^{\mu\nu}\right)\Bigg]\,,
	\end{align}
	The sound modes are given by
	\begin{align*}
		\delta g_{\mathtt v\mathtt v}\,,\,\delta g_{\mathtt v x}\,,\,\delta g_{\mathtt v r}\,,\,\delta g_{rr}\,,\,\delta g_{rx}\,,\,\delta g_{xx}\,.
	\end{align*}
	Imposing radial gauge condition $\delta g_{r\mu}=0$ and from the trace-less condition $g^{\mu\nu}\delta g_{\mu\nu}=0$, the non-redundant modes are only 
	\begin{align*}
		\delta g_{\mathtt v \mathtt v}\,,\,\delta g_{\mathtt v x}\,.
	\end{align*}
	Furthermore, the perturbed equations should be regular at the horizon in the ingoing EF coordinates. Therefore, we expand the modes near the horizon as follows
	\begin{align} \label{regularity}
		\delta g_{\mu\nu}(r)=\delta g_{\mu\nu}^{(0)}+(r-2\sqrt{\CL_\mu})\delta g_{\mu\nu}^{(1)}+\cdots 
	\end{align}
	As described in \cite{Blake:2018leo}, the near horizon expansion of the linearized Einstein's equation $\delta E^{}_{\mathtt v \mathtt v}=0$ becomes degenerate at the special points \eqref{Pole-skipping-point}, admitting an extra ingoing mode and causing coincident zeros and poles in the retarded Green’s function of the energy‑density correlator \cite{Blake:2018leo}. This links the chaos parameters, the Lyapunov exponent $\lambda_L$ and butterfly velocity $v_B$, directly to the analytic structure of thermal two‑point functions, a feature verified in both anisotropic plasma models \cite{Grozdanov:2018kkt} and massive gravity backgrounds \cite{Ceplak:2021efc}. Thus, the Einstein equation, and its generalizations to other bulk fields, provides a universal gravitational origin for quantum chaos in holography. We may expand the Einstein's equations near the horizon using \cref{regularity} for the perturbed sound modes \cref{logitudinal-perturbations}.	Expanding the linearized Einstein equation $\delta E^{}_{\mathtt v\mathtt v}=0$ near the horizon, we find the constraint relation
	%	** Explain the origin of the constraint condition $\delta E^{}_{vv}=0$ **
	\begin{align}
		2 k (1-2 \mu  \CL_\mu)^2 &\left(2 \mu  \CL_\mu \omega-2 i \sqrt{\CL_\mu}+\omega\right)\delta g_{\mathtt v x}^{(0)}\notag\\&+(1+2 \mu  \CL_\mu)\left(k^2 (1-2 \mu  \CL_\mu)^2-2 i \sqrt{\CL_\mu} \omega (1+2 \mu  \CL_\mu)\right)\delta g_{\mathtt v\mathtt v}^{(0)}=0\,.
	\end{align}
	Equating the coefficients of $\delta g_{\mathtt v x}$ and $\delta g_{\mathtt v\mathtt v}$ separately to zero, we may obtain the Pole-skipping points as follows
	\begin{align}
		(\omega_\star,k_\star)&=\left(i\frac{2\sqrt{\CL_\mu}}{1+2\mu\CL_\mu},i\frac{2\sqrt{\CL_\mu}}{1-2\mu\CL_\mu}\right)\notag\\
		&=\left(i\frac{2\pi}{\beta},i\frac{2 \pi}{\sqrt{\beta ^2-8 \pi ^2 \mu }}\right)\,.
	\end{align}
	Therefore, from \cref{Pole-skipping-point} we obtain the same values for the Lyapunov exponent and the butterfly velocity, as computed from the shockwave method and OTOC.

	\subsection{Entanglement wedge method}
\begin{figure}[ht]
		\centering
		\includegraphics[width=0.8\textwidth]{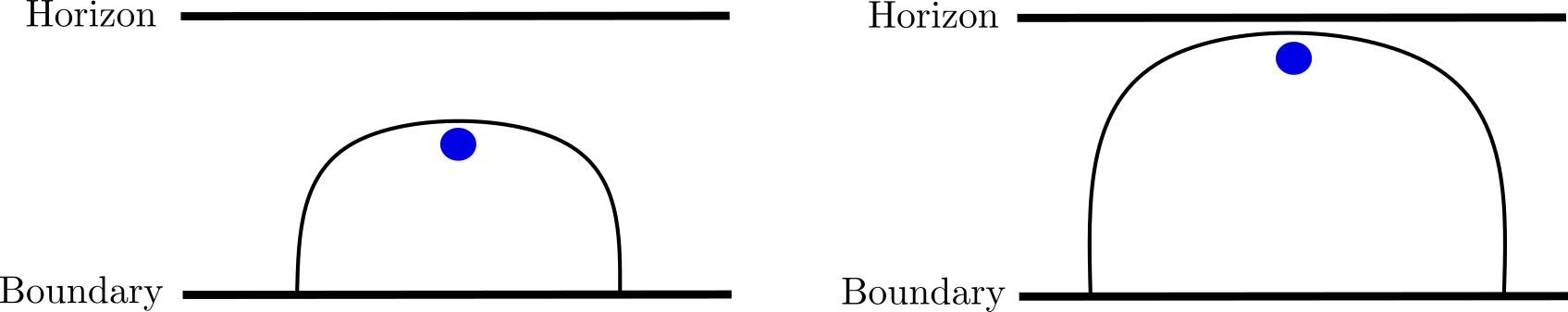}
		\caption{A falling particle being enclosed in the entanglement wedge of a boundary region which grows with the butterfly velocity.}
		\label{fig:EW-method-BTZ}
	\end{figure}

	In this section, we aim to find the butterfly velocity utilizing the entanglement wedge method \cite{Mezei:2016wfz}. We are required to calculate the size of the smallest boundary region for which the falling particle is contained in the entanglement wedge as depicted in \cref{fig:EW-method-BTZ}. At late times, the Ryu-Takayanagi (RT) surface approaches the near-horizon region and exhibits a characteristic profile which propagates outwards at a constant velocity.
	As described in \cite{Ryu:2006bv,Ryu:2006ef}, the holographic entanglement entropy can be obtained in terms of the area of the minimal codimension-two surface (dubbed the RT surface) homologous to the subsystem under consideration. This RT surface is determined by extremizing the area functional
	\begin{align}
		\CA=2\pi\int_\Gamma d^{d-1}y\sqrt{\gamma}\,,\label{Area-functional}
	\end{align}
	where $\gamma$ is determinant of the induced metric on a codimension-two hypersurface $\Gamma$ described by the coordinates $y$. For a constant time slice of the deformed BTZ black hole background in \cref{Deformed-BTZ}, the induced metric on $\Gamma$ is given by
	\begin{align}
		\gamma_{ab}\d x^a\d x^b=\frac{1}{z^2(x)}\left[\frac{z^\prime(x)^2}{1-4\CL_\mu z(x)^2}+\frac{1}{(1-2\mu\CL_\mu)^2}\right]\d x^2\,,
	\end{align}
	where we have parametrized $\Gamma$ as $z=z(x)$ with $z=\frac{1}{r}$. The equation of the RT surface is given by the Euler-Lagranges equation for the area functional \eqref{Area-functional}:
	\begin{align}
		\frac{z(x) \left(4 \CL_\mu z(x)^2-1\right) z''(x)+\left(1-8 \CL_\mu z(x)^2\right) z'(x)^2}{\left(1-4 \CL_\mu z(x)^2\right)^2}=\frac{1}{(1-2 \mu  \CL_\mu)^2}\,.
	\end{align}
	As we work in the near horizon region, we will make the following ansatz for the RT profile \cite{Mezei:2016wfz,Dong:2022ucb}:
	\begin{align}
		z(x)=2\sqrt{\CL_\mu}-\epsilon\,s(x)^2\,.
	\end{align}
where $\epsilon$ is a small parameter and $s(x)$ quantifies the closeness of the RT surface and the horizon. We may Taylor expand the equation of motion around $\epsilon=0$, to obtain the following differential equation for the RT profile\footnote{Alternatively, we may also Taylor expand the induced metric $\gamma$ near the horizon to obtain
	\begin{align}
		\gamma_{ab}\d x^a\d x^b=\left[\frac{4\CL_\mu}{(1-2\mu\CL_\mu)^2}+4\epsilon\sqrt{\CL_\mu}\left(\frac{4\CL_\mu}{(1-2\mu\CL_\mu)^2}s(x)^2+s'(x)^2\right)\right]\,.
		\end{align}
	Taking the variation of the determinant $\sqrt{\gamma}$, we obtain the same differential equation \eqref{DE-s(x)} for the RT profile $s(x)$.}
	\begin{align}
		s''(x)-\frac{4\CL_\mu}{(1-2\mu\CL_\mu)^2}s(x)=0\implies s(x)=e^{\frac{2\sqrt{\CL_\mu}}{1-2\mu\CL_\mu}x}\,.\label{DE-s(x)}
	\end{align}
	At any point of time, we demand that the tip of the RT surface intersects the particle, which is enforced by setting \cite{Mezei:2016wfz}
	\begin{align}
		s(x=0,t)\sim e^{-\frac{2\pi t}{\beta}}\,. \label{DE-s(0)}
	\end{align}
	Using \cref{DE-s(x),DE-s(0)}, the time dependent RT profile for the boundary region is given by
		\begin{align}
		s(x,t)\sim e^{\frac{2\sqrt{\CL_\mu}}{1-2\mu\CL_\mu}x-\frac{2\pi t}{\beta}}\,. 
	\end{align}
	At any given time $t$, there is some size of the boundary region $x=x_\star=\tilde{v}_B t$ such that $s(x_\star,t)\sim \CO(1)$. This gives the minimal size of the boundary region which propagates outward with a characteristic velocity
	\begin{align}
		\tilde{v}_B=\frac{2\pi}{\beta}\frac{1-2\mu\CL_\mu}{2\sqrt{\CL_\mu}}=\sqrt{1-\frac{8\pi^2\mu}{\beta^2}}\,,
	\end{align}
	which clearly aligns with our result for the butterfly velocity in \eqref{vB-BTZ}. Therefore, the entanglement wedge method also corroborates our finding related to the violation of the Mezei-Stanford bound for $\mu<0$.
	%-------------------------------------------------------------------------------
	\section{Rotating BTZ}\label{sec-4}
	Having established the robustness of the three methods in characterizing chaos properties, namely the shockwave method, pole skipping and the entanglement wedge method, for the $\TTbar$ deformed non-rotating BTZ black hole, we now address the issue of characterizing chaos properties of the deformed rotating black holes. As discussed earlier, the entanglement wedge method is only reliable for static geometries in the bulk, it is not suitable for the rotating case and we restrict our attention to the shockwave analysis and the pole skipping phenomena.
	\subsection{Pole skipping}
	As we have already seen, the shockwave and OTOC calculations leads to same characteristics of chaos\footnote{Note that in the rotating case, there are various subtleties related to the wrong choice of affine parametrization and null coordinates due to the presence of an angular momentum of the rotating shockwave which makes the Kruskal extension subtle and correspondingly the shockwave calculations are cumbersome. See for example, \cite{Malvimat:2021itk,Malvimat:2022fhd,Malvimat:2022oue}. In this article, we refrain ourselves from considering rotating shockwaves.}, in this section we will restrict ourselves to the investigation of the pole skipping phenomena in the rotating black hole case. 
	In the Fefferman-graham gauge, the deformed rotating black hole spacetime is given by \cite{Guica:2019nzm}
	\begin{align}
		\d s^2=\frac{\d \rho^2}{4\rho^2}&+\frac{1+4\mu^2\CL_\mu\bar\CL_\mu+\rho\CL_\mu\bar\CL_\mu\left(\rho+8\mu+4\mu^2\CL_\mu\bar\CL_\mu\rho\right)}{\rho\left(1-4\mu^2\CL_\mu\bar\CL_\mu\right)^2}\d Z\,\d \bar Z\notag\\
		&+\frac{(\rho+2\mu)\left(1+2\mu\CL_\mu\bar\CL_\mu\rho\right)}{\rho\left(1-4\mu^2\CL_\mu\bar\CL_\mu\right)^2}\left(\CL_\mu\d Z^2+\bar{\CL}_\mu\d \bar Z^2\right)\,.\label{DrBTZ-FG}
	\end{align}
	Defining the new variables
	\begin{align}
		r^2=\frac{1}{\rho}+\CL_\mu+\bar{\CL}_\mu+\rho\CL_\mu\bar{\CL}_\mu~~,~~Z=x+t~~,~~\bar Z=x-t\,,
	\end{align}
	the above metric may be recast in the following form \cite{Banerjee:2024wtl}
	\begin{align}
		\d s^2=\frac{r^2}{(r^2-r_+^2)(r^2-r_-^2)}\d r^2-(r^2-r_+^2)\frac{(r_+\d t-r_-\d x)^2}{r_h^2\left(1+\frac{r_h^2}{2}\mu\right)^2}+(r^2-r_-^2)\frac{(r_+\d x-r_-\d t)^2}{r_h^2\left(1-\frac{r_h^2}{2}\mu\right)^2}\,,\label{DrBTZ}
	\end{align}
	where we have used the notations
	\begin{align}
		\CL_\mu=\frac{1}{4}(r_+-r_-)^2~~,~~\bar\CL_\mu=\frac{1}{4}(r_++r_-)^2~~,~~r_h^2=r_+^2-r_-^2\,.\label{L-Lb-DrBTZ}
	\end{align}
	This geometry is dual to a $\TTbar$-deformed CFT$_2$ at a finite temperature $\beta^{-1}$ and chemical potential for anugular momentum $\Omega$ given by \cite{Banerjee:2024wtl}
	\begin{align}
		\beta=\frac{\pi}{2}\left(\frac{1}{\sqrt{\CL_\mu}}+\frac{1}{\sqrt{\bar\CL_\mu}}\right)\left(1+2\mu\sqrt{\CL_\mu\bar\CL_\mu}\right)~~,~~\Omega=\frac{\sqrt{\bar\CL_\mu}-\sqrt{\CL_\mu}}{\sqrt{\bar\CL_\mu}+\sqrt{\CL_\mu}}\,.\label{beta-Omega-DrBTZ}
	\end{align}
	For the near horizon analysis, it is convenient to use the comoving coordinates $(r,t,\phi)$ where
	\begin{align}
		\phi=x-\frac{r_-}{r_+}t\equiv x-\Omega t\,,\label{Comoving-coordinate}
	\end{align}
	in which the metric reads
	\begin{align}
		\d s^2=\frac{r^2}{(r^2-r_+^2)(r^2-r_-^2)}\d r^2-\frac{(r^2-r_+^2)}{r_h^2\left(1+\frac{r_h^2}{2}\mu\right)^2}\left(\frac{r_h^2}{r_+}\d t-r_-\d \phi\right)^2+(r^2-r_-^2)\frac{r_+^2\d\phi^2}{r_h^2\left(1-\frac{r_h^2}{2}\mu\right)^2}\,.\label{DrBTZ-comoving}
	\end{align}
	In these coordinates, the angular speed vanishes at the horizon $\Omega_H=0$. 
%	The dual field theory for \eqref{DrBTZ} has the metric $\d s^2=-\d t^2+\d x^2$, while the dual field theory for \eqref{DrBTZ-comoving} lives on the spacetime
%	\begin{align}
%		\d s^2=-\d t^2+(\Omega\d t+\d\phi)^2\,.
%	\end{align}
	These two theories are related by the boost transformation \eqref{Comoving-coordinate}. The momentum variables in the comoving and Schwarzschild coordinates are related by \cite{Liu:2020yaf}
	\begin{align}
		\omega_\textrm{Sch}=\omega_\textrm{cm}+\Omega\,k_\textrm{cm}~~,~~k_\textrm{Sch}=k_\textrm{cm}\,.\label{frequency-transform}
	\end{align}
	In order to analyze ingoing boundary conditions near the horizon, it is customary to use the ingoing Eddington-Finkelstein coordinates. The ingoing EF time $\mathtt v$ can be obtained from the comoving metric as
	\begin{align}
		\mathtt v=t+r_\star~~,~~\d r_\star=\frac{r r_+\left(2+r_h^2\mu\right)}{2r_h\sqrt{r^2-r_-^2}(r^2-r_+^2)}\,,
	\end{align}
	where the tortoise coordinate is given by
	\begin{align}
		r_\star=\frac{r_+\left(1+\frac{r_h^2}{2}\mu\right)}{2r_h^2}\log\left[\frac{\sqrt{r^2-r_-^2}-r_h}{\sqrt{r^2-r_-^2}+r_h}\right]\,.\label{Tortoise-DrBTZ}
	\end{align}
	Therefore, the deformed metric in the ingoing EF coordinates can be written as
	\begin{align}
		\d s^2=&-\frac{(r^2-r_+^2)r_h^2}{r_+^2\left(1+\frac{r_h^2}{2}\mu\right)^2}\d \mathtt v^2+2\frac{r\, r_h}{r_+\sqrt{r^2-r_-^2}}\frac{\d \mathtt v\,\d r}{\left(1+\frac{r_h^2}{2}\mu\right)}+2\frac{r_-}{r_+}(r^2-r_+^2)\frac{\d \mathtt v\,\d\phi}{\left(1+\frac{r_h^2}{2}\mu\right)^2}\notag\\
		&-\frac{2r\,r_-}{r_h\sqrt{r^2-r_-^2}}\frac{\d r\,\d\phi}{\left(1+\frac{r_h^2}{2}\mu\right)}+\frac{4\left(-8r_-^2r_+^2\mu+r^2\left(4+4(r_+^2+r_-^2)\mu+r_h^4\mu^2\right)\right)}{\left(4-r_h^4\mu^2\right)^2}\d\phi^2\,.
	\end{align}
	To proceed, we assume that the angular coordinate $\phi$ is non-compact, namely, we are working in the high temperature limit. We now allow for the perturbations on the longitudinal modes, as in \cref{logitudinal-perturbations} and expand the fields near the (outer) horizon\footnote{Note that, remarkably the Einstein's equations are regular at the inner horizon.}
	\begin{align}
		\delta g_{\mu\nu}=\sum_{n=0}^{\infty}\delta g_{\mu\nu}^{(n)}(r-r_+)^n\,.
	\end{align}
	Then the $vv$-component of the Einstein's equations, in the leading order about the (outer) horizon, results in the following constraint equation
	\begin{align}
		r_+ \left(-k \left(2-\mu  r_h^2\right)^2 \left(-k \left(\mu  r_h^2+2\right)+4 i r_-\right)-2 i r_+ \omega  \left(\mu  (r_--r_+)^2+2\right) \left(\mu  (r_-+r_+)^2+2\right)\right)&\delta g_{\mathtt v\mathtt v}^{(0)}\notag\\
		+2k \left(2-r_h^2\mu\right)^2\left(\omega  r_+\left(2+r_h^2\mu\right)-2 i r_h^2\right)\delta g_{\mathtt v x}^{(0)}&=0\,.
	\end{align}
	Setting the coefficients of $\delta g_{\mathtt vx}\,,\,\delta g_{\mathtt v\mathtt v}$ to zero separately, we obtain the Pole skipping points as follows
	\begin{align}
		(\omega_\star,k_\star)&=\left(\frac{2i r_h^2}{r_+\left(2+r_h^2\mu\right)},\pm\frac{2i(r_+\pm r_-)(2+(r_+\mp r_-)^2\mu)}{4-r_h^4\mu^2}\right)\,,\notag\\
		&=\left(\frac{2\pi i}{\beta},\pm\frac{2\pi i}{\beta(1-\Omega^2)}\left(\Omega\pm\frac{1}{\sqrt{1-\frac{8\pi^2\mu}{\beta_+\beta_-}}}\right)\right)\,,
	\end{align}
	where in the second equality, we have used \cref{L-Lb-DrBTZ,beta-Omega-DrBTZ} to express the horizon radii in terms of the field theoretic parameters $\beta\,,\,\Omega$.
	It should be emphasized that the other components of the Einstein's equations are regular at these points. From the above expression, using \cref{Pole-skipping-point}, we may infer the chaos parameters as follows
	\begin{align}
		\lambda_L=\frac{2\pi}{\beta}~~,~~v_B=\pm\frac{1-\Omega^2}{\Omega\pm\frac{1}{\sqrt{1-\frac{8\pi^2\mu}{\beta_+\beta_-}}}}\,.\label{lambda-vB-comoving}
	\end{align}
	Therefore, in the comoving coordinates, the Lyapunov exponent saturates the MSS bound. We plot the butterfly velocities in \cref{fig:vB-comoving}. Interestingly, the butterfly velocity in the comoving coordinates seem to vanish at some particular value of $\mu$:
	\begin{figure}[ht]
		\centering
		\includegraphics[width=0.5\textwidth]{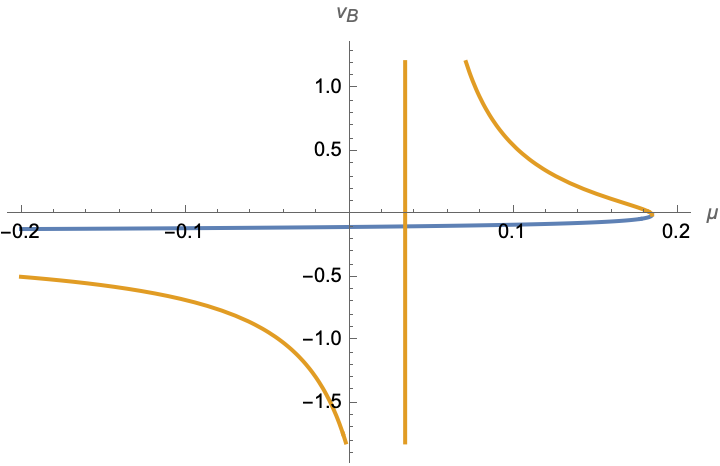}
		\caption{Variation of the butterfly velocity with respect to $\mu$ in the comoving coordinates.}
		\label{fig:vB-comoving}
	\end{figure}
	\begin{align}
		\mu^{}_\textrm{Hagedorn}=\frac{\beta^2(1-\Omega^2)}{8\pi^2}\equiv \frac{\beta_+\beta_-}{8\pi^2}\,.\label{Hagedorn-rBTZ}
	\end{align}
	However, this value corresponds to the spinning analog of the Hagedorn bound \cite{McGough:2016lol,Banerjee:2024wtl}, where the $\TTbar$ deformed energy levels become complex and hence does not correspond to a physically relevant situation. Note that, as $\Omega\to 0$, we find $v_B=\sqrt{1-\frac{8\pi^2\mu}{\beta^2}}$, thereby reproducing the non-rotating result \eqref{vB-BTZ}. Furthermore, as $\mu\to 0$, we reproduce the undeformed results obtained in \cite{Craps:2021bmz}
	\begin{align}
		\lambda_L=\frac{2\pi}{\beta}~~,~~v_B=1\pm \Omega\,.
	\end{align}
	
\begin{figure}[ht]
		\centering
		\includegraphics[width=0.48\textwidth]{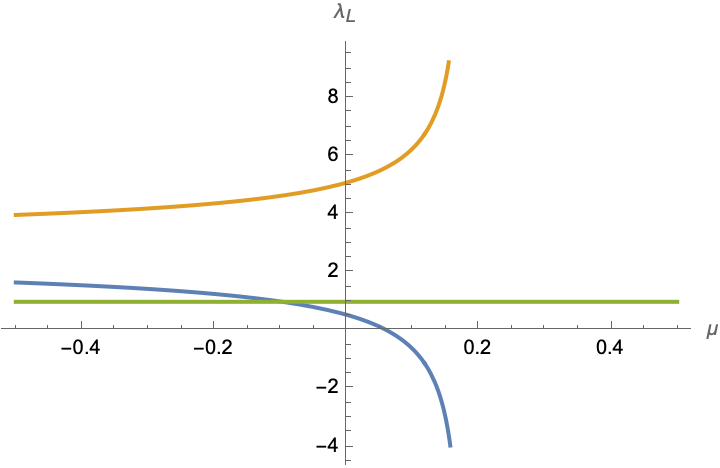}
		\includegraphics[width=0.48\textwidth]{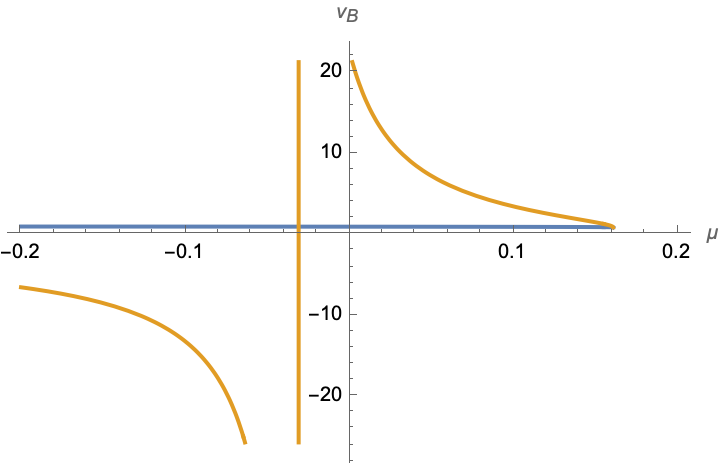}
		\caption{Variation of the Lyapunov exponent and butterfly velocity with respect to $\mu$ in the Schwarzschild coordinates.}
		\label{fig:Lya-vB-Sch}
	\end{figure}

	In the Schwarzschild coordinates, the transformed frequencies are obtained from \cref{frequency-transform}
	\begin{align}
		(\omega_\star,k_\star)=\left(\frac{2 \pi i}{\beta  \left(1-\Omega ^2\right)}\left(1\pm\frac{\Omega }{\sqrt{1-\frac{8 \pi ^2 \mu }{\beta ^2 \left(1-\Omega ^2\right)}}}\right),\pm\frac{2\pi i}{\beta(1-\Omega^2)}\left(\Omega\pm\frac{1}{\sqrt{1-\frac{8\pi^2\mu}{\beta_+\beta_-}}}\right)\right)\,,
	\end{align}
	from which we may read off the chaos parameters as follows
	\begin{align}
		\lambda_L^{\pm}=\frac{2 \pi}{\beta  \left(1-\Omega ^2\right)}\left(1\mp\frac{\Omega }{\sqrt{1-\frac{8 \pi ^2 \mu }{\beta_+\beta_-}}}\right)~~,~~v_B^\pm=\frac{1+\Omega  \sqrt{1-\frac{8 \pi ^2 \mu }{\beta_+\beta_-}}}{\Omega\pm\sqrt{1-\frac{8 \pi ^2 \mu }{\beta_+\beta_-}} }\,.\label{lambda-vB-Sch}
	\end{align}
	Hence, similar to the undeformed case, we find two Lyapunov exponents corresponding to the left and right moving modes in the dual CFT$_2$ with $\TTbar$ deformation. In \cref{fig:Lya-vB-Sch}, we have plotted the two Lyapunov exponents and butterfly velocities with respect to the deformation parameter $\mu$. Note that, once again, if we switch off the rotation of the black hole, the chaos parameters reduce to our earlier findings in \eqref{vB-BTZ}.
	
	We find that $\lambda_L^-$ always seems to violate the chaos bound, similar to the undeformed case and it is necessary to take into account the periodicity of the $\phi$ coordinate to get rid of this pathology, as described in \cite{Jahnke:2019gxr}. However, in the high temperature limit, $\lambda_L^-$ does serve as the upper limit of instantaneous Lyapunov exponent, which may be interpreted as the effective Lyapunov exponent corresponding to the right movers who find themselves immersed in a thermal ensemble at inverse temperature $\beta_-$. Interestingly, in the deformed case, we find that at some particular negative value of $\mu$, given as
	\begin{align}
		\mu_l=-\frac{\beta^2(1-\Omega^2)^2}{8\pi^2\Omega^2}\equiv -\frac{\beta_+^2\beta_-^2}{2\pi^2(\beta_+-\beta_-)^2}\,,
	\end{align} 
	even $\lambda_L^+$ starts to violate the chaos bound.  However, as argued in \cite{Banerjee:2024wtl,Basu:2024enr}, at this value of $\mu$ the $\TTbar$-deformed theory is not well-defined. For example, the spacelike RT surface computing the entanglement entropy of a spatial subsystem in the field theory becomes null at this limit and correspondingly the entanglement entropy is IR divergent. Hence $\mu_l$ serves as a lower bound to the deformation parameter, and correspondingly there is no violation of the MSS bound from the left moving modes. Furthermore, at the Hagedorn temperature \eqref{Hagedorn-rBTZ}, both $\lambda_L^{\pm}$ diverges indicating breakdown of the $\TTbar$ deformed theory. Interestingly, in this limit, the butterfly velocities remains finite, and is equal to the angular speed of the black hole $v_B^\pm(\mu^{}_\textrm{Hagedorn})=\Omega$. This conforms with our earlier findings in the case of non-rotating black holes. Remarkably, in the range
	\begin{align}
		\mu_l<\mu<0\,,
	\end{align}
	the butterfly velocity $v_B^+$ exceeds the Mezei-Stanford bound. As discussed earlier, such behavior is already expected from the non-local nature of the $\TTbar$-deformation.
	%------------------------
	\subsubsection*{Extremal black holes}
	We now briefly comment on the extremal black hole case which corresponds to $r_-\to r_+$ or, alternatively $\Omega\to 1\,,\,\beta\to\infty$. The deformed black hole metric may be obtained from \cref{DrBTZ-FG} by taking the limit $\CL_\mu\to 0$ as follows
	\begin{align}
		\d s^2&=\frac{\d \rho^2}{4\rho^2}+\frac{1}{\rho}\d Z\,\d \bar Z+\left(1+\frac{2\mu}{\rho}\right)\bar\CL_\mu\d \bar Z^2\notag\\
		&=\frac{r^2\d r^2}{(r^2-r_0)^2}+2(r^2-r_0^2)\d t\,\d \phi+\left(2 \mu  r^2 r_0^2+r^2-2 \mu  r_0^4\right)\d\phi^2\,\label{Dr-Ext-BTZ}
	\end{align}
	where, in the second equality, we have made the following change of coordinates
	\begin{align}
		r^2=\frac{1}{\rho}+\bar\CL_\mu~~,~~Z=\phi+2t~~,~~\bar Z=\phi\,.
	\end{align}
	In this case, we find that the Lyapunov exponent and the butterfly velocity both vanish for the right moving modes, whereas for the left moving modes they are independent of the deformation parameter:
	\begin{align}
		\lambda_L^\textrm{ext}=\frac{2\pi}{\beta_\textrm{FT}}~~,~~v_B^\textrm{ext}=1\,,
	\end{align}
	where $\beta_\textrm{FT}=\frac{\pi}{r_0}$ corresponds to the effective Frolov-Thorne temperature. This is surprising, since the deformed metric \eqref{Dr-Ext-BTZ}  is still dependent on $\mu$ in the extremal limit and warrants further investigation.
	%----------------------------------------------------------------------------------------------------
	\subsection{Shockwaves and OTOC}
	In this subsection, we present a brief account of the shockwave solutions in $\TTbar$ deformed rotating BTZ black holes and the corresponding OTOC computations. The maximal Kruskal extension of the deformed rotating BTZ black hole may be obtained utilizing the comoving coordinates \eqref{DrBTZ-comoving} and the tortoise coordinate \eqref{Tortoise-DrBTZ} as follows
	\begin{align}
		u=-e^{-\kappa \mathtt u}\equiv -e^{-\kappa(t-r_\star)}~~,~~v=e^{\kappa \mathtt v}\equiv e^{\kappa(t+r_\star)}\,,
	\end{align}
	with the surface gravity\footnote{The surface gravity in the deformed rotating BTZ geometry \eqref{DrBTZ} may be computed from the Killing vector $\xi=\del_t$ as follows
	\begin{align}
		\kappa^2=-\frac{1}{2}\left(\nabla_{\mu}\xi_\nu\right)\left(\nabla^\mu\xi^\nu\right)\Big|_{r=r_+}\,.
		\end{align}}
	\begin{align}
		\kappa=\frac{r_+^2-r_-^2}{r_+ \left[1+\frac{1}{2} \mu  \left(r_+^2-r_-^2\right)\right]}=\frac{2\pi}{\beta}\,,
	\end{align}
	where in the second equality, we have made use of \cref{L-Lb-DrBTZ,beta-Omega-DrBTZ}. In these coordinates, the deformed black hole metric \eqref{DrBTZ-comoving} takes the form
	\begin{align}
		\d s^2=&-\frac{4 \d u\, \d v}{(1+u v)^2}-\frac{4 r_- (u\d v-v \d u)d\phi}{\left(1+\frac{r_h^2}{2}\mu  \right)(1+u v)^2}+\frac{1}{(1+u v)^2}\left[\frac{4 r_-^2u v}{\left(1+\frac{r_h^2}{2}\mu \right)^2} +\frac{r_+^2(1-u v)^2 }{\left(1-\frac{r_h^2}{2}\mu \right)^2}\right]\d\phi^2\,.
	\end{align}
	We consider perturbing the above solution with the insertion of a localized shockwave
	\begin{align}
		T_{uu}^{\textrm{shock}}=E_0\,e^{\kappa t}\delta(u)\delta(\phi)\,,
	\end{align}
	localized on the horizon $u=0$ at $\phi=0$. Here, $E_0$ is a constant related to the asymptotic energy of the shockwave and $e^{\kappa t}$ represents the blue-shift factor. This has the following effect on the metric \cite{Dray:1985yt,Sfetsos:1994xa}
	\begin{align}
		\d s^2\to \d s^2+\frac{2}{(1+u v)^2}\delta(u)h(\phi)\d u^2\,,
	\end{align}
	where $h(\phi)$ quantifies the shift in the horizon after shockwave insertion.
	Note that the transverse directions are unaffected by the shockwave. Then the $u u$ component of the Einstein equation leads to the following differential equation
	\begin{align}
		&\left(\mu  (r_-^2-r_+^2)-2\right) \left(\mu  (r_-^2-r_+^2)+2\right)^2 \left(h''(\phi ) \left(\mu(  r_-^2-r_+^2)-2\right)+4 r_- h'(\phi )\right)\notag\\
		&+4 h(\phi ) \left(r_-^2-r_+^2\right) \left(\mu ^2 r_-^4-2 \mu  r_-^2 \left(\mu  r_+^2-2\right)+\left(\mu  r_+^2+2\right)^2\right)=8\pi G_N E_0 \,e^{\kappa t}\delta(\phi)\,.\label{DrBTZ-ODE}
	\end{align}
	The general solution to the above equation is given by
	\begin{align}
		h(\phi)&=c_1\,e^{\kappa t} e^{\frac{\left(r_++r_-\right) \left(1+\frac{\mu}{2}\left(r_+-r_-\right)^2\right)}{1-\frac{\mu ^2}{4}\left(r_+^2-r_-^2\right)^2}}+c_2\,e^{\kappa t} e^{-\frac{\left(r_+-r_-\right) \left(1+\frac{\mu}{2}\left(r_++r_-\right)^2\right)}{1-\frac{\mu ^2}{4}\left(r_+^2-r_-^2\right)^2}}\,,\notag\\
		&=c_1\exp\left[\frac{2\pi}{\beta}\left(t-\frac{\Omega +\frac{1}{\sqrt{1-\frac{8 \pi ^2 \mu }{\beta_+\beta_-}}}}{1-\Omega^2}\phi\right)\right]+c_2\exp\left[\frac{2\pi}{\beta}\left(t+\frac{\Omega -\frac{1}{\sqrt{1-\frac{8 \pi ^2 \mu }{\beta_+\beta_-}}}}{1-\Omega^2}\phi\right)\right]\,,
	\end{align}
	where we have absorbed the step functions inside the coefficients $c_{1,2}$. Remarkably, identifying the shockwave profile with the OTOC, we find that the shockwave analysis leads to the same chaos parameters as obtained from the pole skipping analysis in \cref{lambda-vB-comoving}.
	%-----------------------------------------------------------------------------------------
	\subsubsection{Periodicity of $\phi$ } 
	As discussed earlier, the above analysis is only true for the case in which the angular direction $\phi$ is non-compact, that is in the high temperature limit. As described in \cite{Jahnke:2019gxr,Mezei:2019dfv}, for finite temperatures, the correct interpretation of the behavior of OTOC requires taking into account the periodicity of the function $h(\phi)$. In the following, we will show that the above behavior is reproduced in the high temperature limit of the correct Lyapunov exponent.
	
\begin{figure}[ht]
		\centering
		\includegraphics[width=0.5\textwidth]{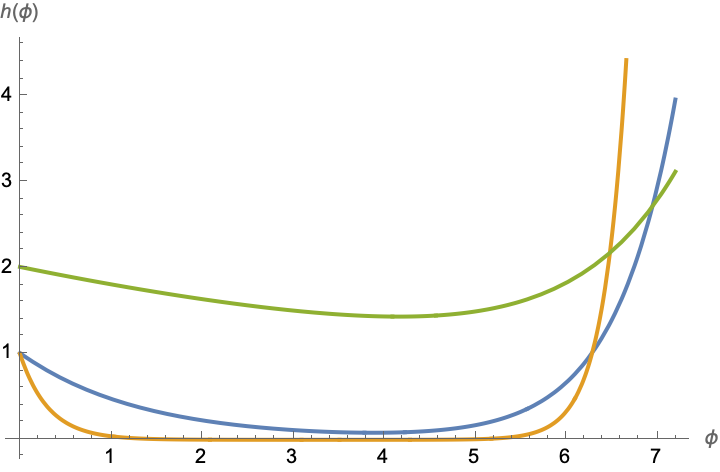}
		\caption{The shock profile in deformed rotating BTZ geometry. We have set $\Omega=\frac{1}{3}\,,\,\beta=2\pi$ and $\mu=0\,\textrm{(blue)}\,,\,0.4\,\textrm{(yellow)}\,,\,-1.95\,\textrm{(green)}$.}
		\label{fig:h-phi}
	\end{figure}

\begin{figure}[ht]
		\centering
		\includegraphics[width=0.49\textwidth]{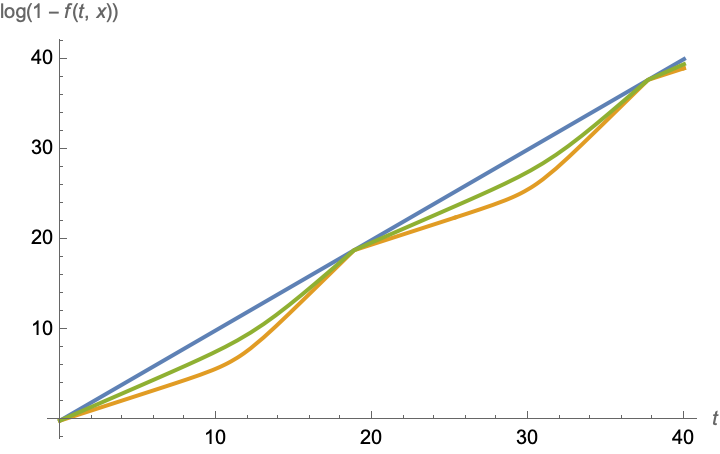}
		\includegraphics[width=0.49\textwidth]{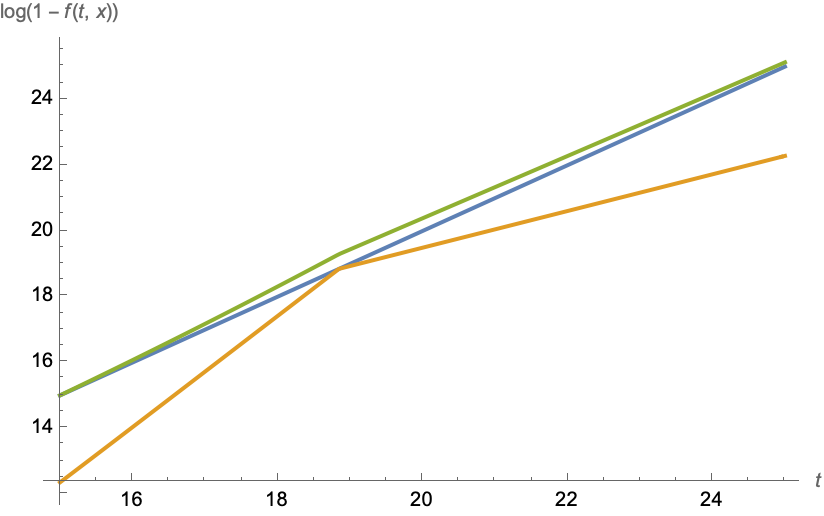}
		\caption{Modulation of the OTOC for different deformation parameters. In the right panel, for $\mu<0$, the modulation grows beyond the chaos bound, shown by the blue curve. The average Lyapunov exponent still saturates the bound $\overline{\lambda}_L=\frac{2\pi}{\beta}$. We have set $\beta=2\pi\,,\,\Omega=\frac{1}{3}$ and [left panel] $\mu=0\,\textrm{(yellow)}\,,\,0.25\,\textrm{(green)}$; [right panel] $\mu=0\,\textrm{(yellow)}\,,\,-1.5\,\textrm{(green)}$.}
		\label{fig:OTOC-modulation}
	\end{figure}

	We impose the periodicity by replacing $h(\phi)\to h(\phi\,\,\textrm{mod} \,2\pi)$. This may be achieved by replacing $\delta(\phi)$ with $\sum_{n=0}^{\infty}\delta(\phi-2\pi n)$ in \cref{DrBTZ-ODE} (see e.g. \cite{Jahnke:2019gxr,Mezei:2019dfv}). Therefore, restricting the domain of $\phi$ as $\phi\in[0,2\pi)$, we obtain the solution
	\begin{align}
		h(\phi)&=e^{\kappa t}\left[\sum_{n=-\infty}^{0}e^{-\frac{2\pi(\phi-2\pi n)}{\beta v_B^+}}+\sum_{n=1}^{\infty}e^{\frac{2\pi(\phi-2\pi n)}{\beta v_B^-}}\right]\,,\notag\\
		&=e^{\kappa t}\left[\frac{e^{-\frac{2\pi}{\beta v_B^+}\phi\,(\textrm{mod}\,2\pi)}}{1-e^{-\frac{4\pi^2}{\beta v_B^+}}}+\frac{e^{\frac{2\pi}{\beta v_B^-}\phi\,(\textrm{mod}\,2\pi)}}{e^{\frac{4\pi^2}{\beta v_B^-}}-1}\right]\,,
	\end{align}
	where we have denoted
	\begin{align}
		v_B^{\pm}=\pm\frac{1-\Omega^2}{\Omega\pm\frac{1}{\sqrt{1-\frac{8\pi^2\mu}{\beta_+\beta_-}}}}\,.
	\end{align}
	We have plotted the function $h(\phi)$ for different values of $\mu$ in \cref{fig:h-phi}.
	Now the OTOC may be computed using the following expression\footnote{Note that our definition of the OTOC misses a factor of $e^{\kappa t}$ as compared to \cite{Jahnke:2019gxr}, since in our conventions $h(\phi)$ already carries the same prefactor.} \cite{Jahnke:2019gxr}
	\begin{align}
		f(t,x)=1-\epsilon_{VW}h(\Omega t-x)\,,
	\end{align}
	which is governed by two sets of exponents and the corresponding velocities. In \cref{fig:OTOC-modulation}, we plot the OTOC with respect to time keeping $x$ constant. The function grows as $e^{2\pi t/\beta}$ with a periodic modulation for different values of the deformation parameter. Notice the unusual feature for $\mu_l<\mu<0$, where the modulation grows beyond the chaos bound, as compared to the undeformed case discussed in \cite{Mezei:2019dfv}. However, we find that the average Lyapunov exponent still saturates the chaos bound
	\begin{align}
		\overline{\lambda}_L=\frac{2\pi}{\beta}\,.
	\end{align}
	%-----------------------------------------------------------------------------------------
	\subsubsection{Instantaneous Lyapunov exponent} 
\begin{figure}[ht]
		\centering
		\includegraphics[width=0.49\textwidth]{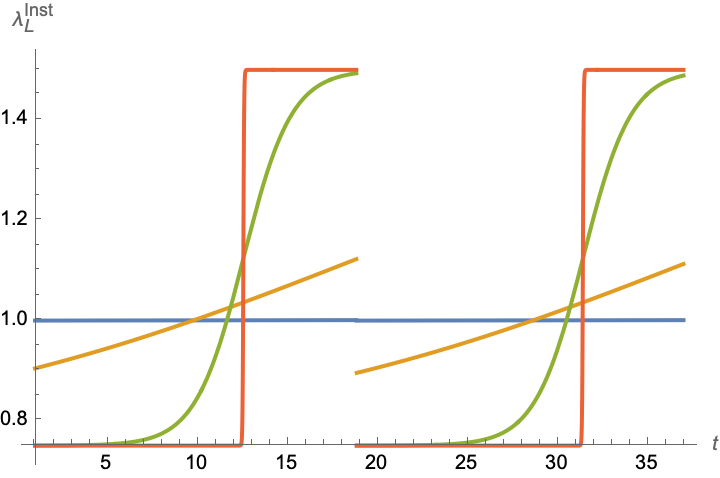}
		\includegraphics[width=0.49\textwidth]{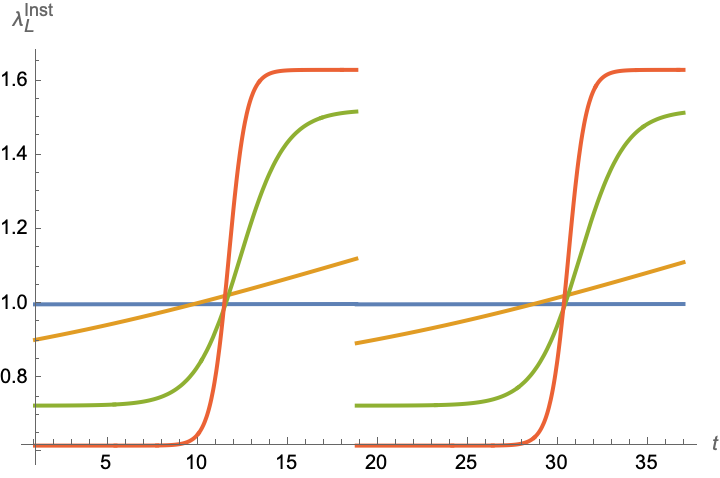}
		\caption{Instantaneous Lyapunov exponent with respect to time. [Left panel] $\mu=0$, [right panel] $\mu=0.05$. We have set $\Omega=\frac{1}{3}$ and plotted $\lambda_\textrm{inst}$ for different values of $\beta=0,\,2\pi,\,20\pi,\,\infty$ (left panel) and $\beta=\pi,\,2\pi,\,20\pi,\,\lesssim \beta_\textrm{Hagedorn}$ (right panel).}
		\label{fig:instantaneous-lambda}
	\end{figure}

	In \cref{fig:instantaneous-lambda}, we plot the instantaneous Lyapunov exponent \cite{Mezei:2019dfv}, defined as follows
	\begin{align}
		\lambda_\textrm{inst}(t)=\frac{\left|\del_t f(t,0)\right|}{1-f(t,0)}=\frac{2\pi}{\beta}+\frac{\del_t h(\Omega t)}{h(\Omega t)}\,.
	\end{align}
	Notice that in the presence of deformation, we cannot go to arbitrarily high temperatures beyond the Hagedorn bound for $\mu>0$ and observe the step function behavior of $\lambda_\textrm{inst}$. However as seen in the right panel of  \cref{fig:instantaneous-lambda}, for small values of $\mu$, the high temperature behavior (below the Hagedorn temperature $\beta_\textrm{Hagedorn}=\sqrt{\frac{8\pi^2\mu}{1-\Omega^2}}$) shows a sharp transition between the values $\lambda_L^\pm$ reported in \eqref{lambda-vB-Sch}.
	
	Furthermore, it is easy to verify that
	\begin{align}
		\left|\frac{\left(\del_t+\Omega\del_x\right)f(t,x)}{1-f(t,x)}\right|=\frac{2\pi}{\beta}\,.
	\end{align}
	Therefore, the OTOC saturates the modified chaos bound for the ``Hamiltonian" $\beta(H+\Omega \,J)$ in the rotating ensemble, even in the presence of the non-local $\TTbar$ deformation \cite{Mezei:2019dfv}. 
	
	%%%%%%%%%%%%%%%%%%%%%%%%%%%%%%%%%%%%%%%%%%%%%%%%%%%%%%%%%%%%%%%%%%%%%%%%

	\section{Discussions}\label{sec-5}
	
	To summarize, we have performed a detailed investigation of quantum chaos in the context of holographic $\TTbar$ deformed theories by focusing on both non-rotating and rotating BTZ black holes in the bulk dual described in \cite{Guica:2019nzm}. In particular, we considered three complementary methods such as shockwave geometries, pole-skipping phenomena, and the entanglement wedge approach to characterize the chaos parameters, namely the Lyapunov exponent ($\lambda_L$) and the butterfly velocity ($v_B$). In this context, we first consider the deformed non‐rotating BTZ black hole geometry which can be written in Kruskal coordinates. Later, we introduce small perturbation to the corresponding geometry by adding few particles of very small energy from the left asymptotic boundary. This perturbation, while infinitesimal at early times, becomes an enormous blueshift as it approaches the horizon and backreacts to produce a gravitational shockwave. Utilizing the matching conditions across the null surface in the corresponding geometry, we obtained the backreacted metric in closed form which is valid upto leading order in small energy.
	
	We performed a holographic computation of OTOC by determining the geodesic lengths in a non-rotating deformed BTZ black hole geometry in Kruskal coordinates for both spherically symmetric and localized shockwave scenarios. Finally, we verified the chaos parameters from two different methods namely pole skipping and entanglement wedge method which are consistent with our corresponding results. In addition, we extend our computations of the holographic chaos for the deformed rotating BTZ black hole case. In this regards, we mainly focused on shockwave method for the study of OTOC and the pole skipping. We find that our results of the chaos parameters are consistent with the earlier literature \cite{Jahnke:2019gxr,Malvimat:2021itk} when the deformation $\mu$ is zero.

	Interestingly for the non-rotating BTZ black hole, we observed that the Lyapunov exponent ($\lambda_L = \frac{2\pi}{\beta}$), obtained from pole-skipping points and OTOCs in the shockwave geometry is independent of the deformation parameter $\mu$ and saturates the Maldacena-Shenker-Stanford bound described in \cite{Maldacena:2015waa}. In the rotating case, the comoving coordinates yield a Lyapunov exponent that saturates the MSS bound. However in Schwarzschild coordinates, the left- and right-moving modes exhibit distinct exponents, with $\lambda_L^-$ potentially violating the bound unless periodicity on $\phi$ coordinate is enforced. This behavior aligns with prior studies of undeformed rotating BTZ black holes \citep{Jahnke:2019gxr,Malvimat:2021itk}, but the $\TTbar$ deformation introduces a critical threshold at $\mu_l$, beyond which the theory becomes ill-defined, describing the non-local nature of the deformation.

The butterfly velocity, however, exhibits a striking dependence on the $\TTbar$ deformation. For the non-rotating BTZ black hole, we find butterfly velocity as $v_B = \sqrt{1 - \frac{8\pi^2 \mu}{\beta^2}}$, which is obtained consistently from shockwave analysis, pole-skipping, and the entanglement wedge method. This result indicates a violation of the Mezei-Stanford bound, $v_B \leq 1$ for $d=2$, when the deformation parameter is negative ($\mu < 0$). It can be understood from the entanglement wedge method that the boundary region enclosing a falling particle propagates with the same $v_B$, which exceeds unity for negative $\mu$. In the rotating case, the butterfly velocity in Schwarzschild coordinates further reveals a regime ($\mu_l < \mu < 0$) where $v_B^+$ exceeds the Mezei-Stanford bound, while at the Hagedorn bound ($\mu = \mu_{\text{Hagedorn}}$), $v_B^\pm$ reduces to the angular speed of the black hole. These violations suggest that the non-local effects of $\TTbar$ deformation allow chaos to propagate faster than expected in two-derivative gravity.%, challenging causality constraints in the bulk.

We have provided a field theoretic analysis of the OTOC in \cref{OTOC_Field} for the $\TTbar$ deformed theories following \cite{He:2019vzf}, where we find consistent result of the OTOC up to linear order in $\mu$ corroborating the holographic correspondence in the deformed setting. Additionally, we show that similar arguments hold from an analysis of the induced metric at the asymptotic boundary. Furthermore, our results are consistent with the cut-off prescription \cite{McGough:2016lol} discussed in \cref{AppB}, wherein the boundary theory resides at a finite radial cutoff. The rescaled spatial coordinate in the cut-off picture matches that of the mixed boundary condition approach, ensuring that chaos parameters computed in both frameworks are consistent. This equivalence strengthens the holographic interpretation of $\TTbar$ deformation as a geometric modification of the bulk, either via a finite cutoff or mixed boundary conditions at the asymptotic boundary \citep{McGough:2016lol, Guica:2019nzm}.  In \cref{AppC}, we investigate the late time behavior of the mutual information for the case of spherical shock in the deformed non-rotating BTZ geometry and obtained the scrambling time which exhibits non-trivial dependence on the deformation parameter $\mu$.

		For future directions, it would be interesting to explore the implications of our results for other holographic probes, such as   complexity, in $\TTbar$ deformed theories. A crucial outstanding issue in this context would be a non-perturbative analysis of the Krylov complexity \cite{Parker:2018yvk,Dymarsky:2021bjq,Kundu:2023hbk} in $\TTbar$-deformed AdS$_3$ geometries utilizing the holographic auto-correlation function as well as the momentum-Krylov complexity correspondence \cite{Fan:2024iop}, extending the perturbative analysis in \cite{Chattopadhyay:2024pdj}. The behavior of the butterfly velocity near the Hagedorn bound requires further investigation, particularly in relation to the stability of the deformed CFT$_2$. 
		
		%Extending our analysis to higher-dimensional AdS spacetimes or other irrelevant deformations could clarify whether the violation of the Mezei-Stanford bound is unique to $\TTbar$ deformation in two dimensions. 

	\begin{comment}
		content...

	\begin{itemize}
		\item The bound on the butterfly velocity was obtained through the entanglement wedge method by Stanford and Mezei. Interestingly, our calculations for the deformed bulk as well as the cut-off geometry through the entanglement wedge method suggests a violation of this bound, which substantiates our earlier calculations.
		\item It is easy to check that these three methods, applied to the cut-off AdS holography, lead to the same results for the Lyapunov exponent as well as the butterfly velocity (cf. \cref{AppB}).
	\end{itemize}
		\end{comment}
		
		\section*{Acknowledgment}
		The authors are supported by the NSFC Grant No. 12447108 and the Shing-Tung Yau Center of Southeast University. The authors are grateful to Yunfeng Jiang, Suchetan Das and Vinay Malvimat for useful suggestions and comments on the draft.
		
		%%%%%%%%%%%%%%%%%%%%%%%%%%%%%%%%%%%%%%%%%%%%%%%%%%%%%%%%%%%%%%%%%%%%%
	\appendix
	\section{OTOC in $\TTbar$ deformed CFT$_2$}\label{OTOC_Field}
	\subsection{Conformal perturbation theory}
	In this appendix, we investigate the OTOC between pairs of operators in the $\TTbar$ deformed CFT$_2$ at a finite temperature, defined on a cylinder $\CM$ with circumference $\beta$: 
	\begin{align}
		\frac{\left<W(t)VW(t)V\right>_\beta}{\left<W(t)W(t)\right>_\beta\, \left<VV\right>_\beta}\label{OTOC}
	\end{align}
	in order to investigate the chaotic properties perturbatively. These computations mostly follow \cite{He:2019vzf} and is included to make the manuscript self-contained. Note that the action functional for the $\TTbar$ deformed CFT$_2$ in \cref{TTbar-defn} may be expanded for small deformation parameter $\mu$ as follows
	\begin{align}
		\CS^{[\mu]}=\CS_\textrm{CFT}-\frac{\mu}{\pi^2}\int_\CM d^2x\,T\bar T\,,\label{perturbative-action}
	\end{align}
	where we have assumed a flat background metric $\gamma_{ab}\d x^a\d x^b=\d w\,\d\bar{w}$, and denoted\footnote{Note that the trace $T_{w\bar{w}}$ has been dropped assuming that the seed theory is a CFT.}
	\begin{align}
		T:=2\pi T_{ww}~~,~~\bar{T}=2\pi T_{\bar{w}\bar{w}}\,.
	\end{align}
    According to \eqref{perturbative-action}, any correlation function in the deformed theory may be computed perturbatively as follows
    \begin{align}
    	\left<\CO_1(w_1,\bar w_1)\cdots\right>^{[\mu]}&=\int\CD\phi\left(\CO_1(w_1,\bar w_1)\cdots\right)\,e^{-\CS^{[\mu]}[\phi]}\notag\\
    	&=\int\CD\phi\left(\CO_1(w_1,\bar w_1)\cdots\right)\,e^{-\CS_\textrm{CFT}+\frac{\mu}{\pi^2}\int_\CM\d^2x\,T\bar T}\notag\\
    	&\approx\int\CD\phi\left(\CO_1(w_1,\bar w_1)\cdots\right)\,e^{-\CS_\textrm{CFT}}\left(1+\frac{\mu}{\pi^2}\int_\CM\d^2x\,T\bar T\CO(\mu^2)\right)\notag\\
    	&=\left<\CO_1(w_1,\bar w_1)\cdots\right>^{[0]}+\frac{\mu}{\pi^2}\int d^2x \left<T(w)\bar T(\bar{w})\CO_1(w_1,\bar w_1)\cdots\right>^{[0]}+\CO(\mu^2)
    \end{align}
    where $\phi$ collectively denotes the field content of the CFT and the superscripts on the correlators denote whether they are evaluated in the (un)deformed theory. Now the leading correction to the four-point OTOC in \cref{OTOC} may be obtained as follows
    \begin{align}
    	&\frac{\left<W(w_1,\bar w_1)W(w_2,\bar w_2)V(w_3,\bar w_3)V(w_4,\bar w_4)\right>_\beta}{\left<W(w_1,\bar w_1)W(w_2,\bar w_2)\right>_\beta\, \left<V(w_3,\bar w_3)V(w_4,\bar w_4)\right>_\beta}\notag\\
    	\times&\Bigg(1+\frac{\mu}{\pi^2}\int_\CM\d^2x\frac{\left<T(w)\bar{T}(w)W(w_1,\bar w_1)W(w_2,\bar w_2)V(w_3,\bar w_3)V(w_4,\bar w_4)\right>_\beta}{\left<W(w_1,\bar w_1)W(w_2,\bar w_2)V(w_3,\bar w_3)V(w_4,\bar w_4)\right>_\beta}\notag\\&-\frac{\mu}{\pi^2}\int_\CM\d^2x\frac{\left<T(w)\bar{T}(w)W(w_1,\bar w_1)W(w_2,\bar w_2)\right>_\beta}{\left<W(w_1,\bar w_1)W(w_2,\bar w_2)\right>_\beta}-\frac{\mu}{\pi^2}\int_\CM\d^2x\frac{\left<T(w)\bar{T}(w)V(w_3,\bar w_3)V(w_4,\bar w_4)\right>_\beta}{\left<V(w_3,\bar w_3)V(w_4,\bar w_4)\right>_\beta}\Bigg)\label{A5}
    \end{align} 
    where we have dropped the superscripts for brevity. We map the desired correlators onto the complex plane $\mathbb{C}$ using the conformation transformations
    \begin{align}
    	z=e^{\frac{2\pi w}{\beta}}~~~,~~~\bar z=e^{\frac{2\pi \bar w}{\beta}}
    \end{align}
    Under this transformation, the stress tensor components transform as follows
    \begin{align}
    	T(w)=\left(\frac{2\pi z}{\beta}\right)^2T(z)-\frac{\pi^2c}{6\beta^2}~~,~~\bar T(w)=\left(\frac{2\pi \bar z}{\beta}\right)^2\bar T(z)-\frac{\pi^2c}{6\beta^2}\,,
    \end{align}
    where $c$ is the central charge of the undeformed CFT$_2$. Now the correlation functions with stress tensor insertions may be computed using the conformal Ward identities as follows \cite{Jeong:2019ylz,Basu:2023bov,Basu:2024bal}:
    \begin{align}
    	&\frac{\left<T(w)\bar{T}(w)W(w_1,\bar w_1)W(w_2,\bar w_2)V(w_3,\bar w_3)V(w_4,\bar w_4)\right>_\beta}{\left<W(w_1,\bar w_1)W(w_2,\bar w_2)V(w_3,\bar w_3)V(w_4,\bar w_4)\right>_\beta}\notag\\
    	=&\frac{1}{\left<W(w_1,\bar w_1)W(w_2,\bar w_2)V(w_3,\bar w_3)V(w_4,\bar w_4)\right>_\mathbb{C}}\left[-\frac{\pi^2c}{6\beta^2}+\left(\frac{2\pi z}{\beta}\right)^2\sum_{j=1}^{4}\left(\frac{h_j}{(z-z_j)^2}+\frac{1}{z-z_j}\del_j\right)\right]\notag\\
    	&\times \left[-\frac{\pi^2c}{6\beta^2}+\left(\frac{2\pi \bar z}{\beta}\right)^2\sum_{j=1}^{4}\left(\frac{\bar h_j}{(\bar z-\bar z_j)^2}+\frac{1}{\bar z-\bar z_j}\bar\del_j\right)\right]\left<W(w_1,\bar w_1)W(w_2,\bar w_2)V(w_3,\bar w_3)V(w_4,\bar w_4)\right>_\mathbb{C}\label{A8}
    \end{align}
    The four-point function on the complex plane may be expanded in terms of conformal blacks, and in the large central charge limit with $\frac{h_w}{c}$ fixed and $1\ll h_v\ll c$, the dominant contribution comes from the vacuum block given by \cite{Fitzpatrick:2014vua}
    \begin{align}
    	\CF_0(\eta)=\left(\frac{\eta}{1-(1-\eta)^{1-12 h_w/c}}\right)^{2h_v}~~,~~\eta=\frac{z_{12}z_{34}}{z_{13}z_{24}}
    \end{align}
    Near $\eta\sim 0$, the block simplifies to
    \begin{align}
    	\CF_0(\eta)\approx \left(\frac{1}{1-\frac{24\pi i h_w}{c\eta}}\right)^{2h_v}~~,~~\bar\CF_0(\bar\eta)\sim 1\,.
    \end{align}
    The left hand side of \eqref{A8} becomes
    \begin{align}
    	\left(-\frac{\pi^2c}{6\beta^2}\right)^2&-\frac{\pi^2c}{6\beta^2}\left(\frac{2\pi \bar z}{\beta}\right)^2\sum_{j=1}^4\frac{\bar h_j}{(\bar z-\bar z_j)^2}-\frac{\pi^2c}{6\beta^2}\left(\frac{2\pi z}{\beta}\right)^2\sum_{j=1}^4\left(\frac{h_j}{(z-z_j)^2}+\frac{\del_j\log\CF_0(\eta)}{z-z_j}\right)\notag\\
    	&+\left(\frac{2\pi z}{\beta}\right)^2\left(\frac{2\pi \bar z}{\beta}\right)^2\sum_{i=1}^4\left(\frac{h_i}{(z-z_i)^2}+\frac{\del_i\log\CF_0(\eta)}{z-z_i}\right)\sum_{j=1}^4\frac{\bar h_j}{(\bar z-\bar z_j)^2}\label{A11}
    \end{align}
    Following \cite{Roberts:2014ifa,Perlmutter:2016pkf}, to compute the leading correction to the OTOC, we now place the operators as 
    \begin{align}
    	&z_1=e^{\frac{2\pi}{\beta}i\epsilon_1}~,&\bar z_1=e^{-\frac{2\pi}{\beta}i\epsilon_1}\notag\\
    	&z_2=e^{\frac{2\pi}{\beta}i\epsilon_2}~,&\bar z_2=e^{-\frac{2\pi}{\beta}i\epsilon_2}\notag\\
    	&z_3=e^{\frac{2\pi}{\beta}(t-x+i\epsilon_3)}~,&\bar z_3=e^{\frac{2\pi}{\beta}(-t-x-i\epsilon_3)}\notag\\
    	&z_4=e^{\frac{2\pi}{\beta}(t-x+i\epsilon_4)}~,&\bar z_4=e^{\frac{2\pi}{\beta}(-t-x-i\epsilon_4)}
    \end{align}
    where without loss of generality, we may choose \cite{Perlmutter:2016pkf}
    \begin{align}
    	\epsilon_2=\epsilon_1+\frac{\beta}{2}~~,~~\epsilon_4=\epsilon_3+\frac{\beta}{2}~~,~~\epsilon_1=0\,.
    \end{align}
    Upon simplifying \eqref{A11}, adding the contributions from the two-point functions in \cref{A5} and subsequently performing the integrals utilizing the method elaborated in \cite{Chen:2018eqk,Jeong:2019ylz}, we may obtain the OTOC as follows
    \begin{align}
    		\frac{\left<W(t)VW(t)V\right>_\beta}{\left<W(t)W(t)\right>_\beta\, \left<VV\right>_\beta}\left(1+\mu f_1(x)+\mu f_2(x)e^{\frac{2\pi t}{\beta}}+\cdots\right)
    \end{align}
    where $f_{1,2}(x)$ are complicated functions of $x$. Note that, similar to \cite{He:2019vzf}, we have neglected a non-dynamical divergent piece originating from the integral of the constant term in \cref{A11} which only depends on the cut-off. In the above expression, the leading late time behavior is still governed by the factor $e^{\frac{2\pi t}{\beta}}$ and hence the Lyapunov exponent is undeformed. One may also perform similar computations in the presence of a finite angular momentum utilizing the techniques in \cite{Basu:2024enr} to map the twisted cylinder with periodic identifications on both directions into the thermal cylinder with only a temporal identification.
    %----------------------------------------------------------------------------------
    \subsection{From induced metric at the asymptotic boundary}\label{AppA2}
    Alternatively, we may perform the computations of OTOC in the dual field theory by noting that the induced metric on the asymptotic boundary of \eqref{Deformed-BTZ} is given by
    \begin{align}
    	\d s^2_\textrm{bdy}=-\frac{\d t^2}{(1+2\mu \CL_\mu)^2}+\frac{\d x^2}{(1-2\mu\CL_\mu)^2}\equiv \frac{1}{(1+2\mu \CL_\mu)^2}\left(-\d t^2+\d\hat{x}^2\right)
    \end{align}
    which is conformally flat with the rescaled spatial coordinate
    \begin{align}
    	\hat{x}=\frac{1+2\mu\CL_\mu}{1-2\mu\CL_\mu}x
    \end{align}
    Therefore, the undeformed result for the OTOC computed in \cite{Roberts:2014ifa} is easily translated to the deformed case as follows\footnote{We follow the conventions in \cite{Roberts:2014ifa}. Note that the conformal factors get canceled in the numerator and denominator.}
    \begin{align}
    	\frac{\left<W(t)VW(t)V\right>_\beta}{\left<W(t)W(t)\right>_\beta\, \left<VV\right>_\beta}&\approx \left[\frac{1}{1+\frac{24\pi i h_w}{\epsilon_{12}^\star\epsilon_{34}}e^{\frac{2\pi}{\beta}(t-t_\star-\hat{x})}}\right]^{2h_v}\notag\\
    	&=\left[{1+\frac{24\pi i h_w}{\epsilon_{12}^\star\epsilon_{34}}e^{\frac{2\pi}{\beta}\left(t-t_\star-\frac{\hat{x}}{\sqrt{1-\frac{8\pi^2\mu}{\beta^2}}}\right)}}\right]^{-2h_v}
    \end{align}
    where we have utilized \cref{BTZ-temperature}. Remarkably, the above expression correctly reproduces both the Lyapunov exponent and the butterfly velocity as obtained from the bulk computations. In the presence of a finite chemical potential, a similar analysis of the induced metric on the asymptotic boundary of \eqref{DrBTZ-comoving}, it is possible to reproduce our results following the analysis in \cite{Craps:2021bmz}.
    
    \section{Connection with the cut-off prescription}\label{AppB}
    Consider the undeformed (non-rotating) BTZ black hole with metric
    \begin{align}
    	\d s^2=-(r^2-r_h^2)\d t^2+\frac{\d r^2}{r^2-r_h^2}+r^2\d x^2\,.
    \end{align}
    The induced metric at the cut-off surface $r=r_c$ is given by
    \begin{align}
    	\d s^2=(r_c^2-r_h^2)\left(-\d t^2+\frac{\d x^2}{1-\frac{r_h^2}{r_c^2}}\right):=(r_c^2-r_h^2)\left(-\d t^2+\d\tilde{x}^2\right)~~,~~r_h=\frac{2\pi}{\beta}
    \end{align}
    where the conformal spatial coordinate is given as \cite{Chen:2018eqk,Jeong:2019ylz}
    \begin{align}
    	\tilde{x}=x\left(1-\frac{r_h^2}{r_c^2}\right)^{-1/2}\label{cut-off-rescale}
    \end{align}
    Recall that in the cut-off prescription, the cut-off radius in the bulk is related to the deformation parameter $\mu$ in the boundary theory, in our conventions, as follows \cite{Chen:2018eqk,Jeong:2019ylz}
    \begin{align}
    	r_c^2=\frac{1}{2\mu}\,.
    \end{align}
    Therefore the rescaled spatial coordinate in \eqref{cut-off-rescale} is given by
    \begin{align}
    	\tilde{x}=\frac{x}{\sqrt{1-\frac{8\pi^2\mu}{\beta^2}}}
    \end{align}
    which is identical to $\hat{x}$ in the MBC perspective. Therefore, all our bulk computations using the MBC picture may also be consistently reproduced from the cut-off picture. Note that, although the conformal factors are different in the two perspectives, we may ignore this discrepancy at least for the determination of chaos parameters.
    
    \section{Mutual Information}\label{AppC}
    In this appendix, we investigate the late time behavior of the mutual information between two subsystems in the shockwave geometries corresponding to $\TTbar$ deformed AdS$_3$. For simplicity, we restrict our analysis to the spherical shock in the deformed non-rotating BTZ black hole \eqref{Deformed-BTZ}.
    Consider two equal size subsystems $A=\left[-\frac{\ell}{2},\frac{\ell}{2}\right]$ at $t_L=0$ on the left asymptotic boundary and $B=\left[-\frac{\ell}{2},\frac{\ell}{2}\right]$ at $t_R=0$ on the right asymptotic boundary.
    The mutual information between regions $A$ and $B$ is expressed as:
    \begin{align}
    	I(A:B) = S_A + S_B - S_{AB},
    \end{align}
    where the entanglement entropies of subsystems $A$ and $B$ are given by the well-known result independent of the shockwave \cite{Shenker:2013pqa}
    \begin{align}
    	S_A=S_B&=\frac{1}{2G_N}\log\left[\frac{r_\infty}{\sqrt{L}}\sinh\frac{\sqrt{\CL_\mu}\ell}{1-2\mu\CL_\mu}\right]\,,\notag\\
    	&=\frac{1}{2G_N}\log\left[\frac{\beta-\sqrt{\beta^2-8\pi^2\mu}}{4\pi\mu\,\epsilon_c}\sinh\frac{\pi\ell}{\sqrt{\beta^2-8\pi^2\mu}}\right]\,,
    \end{align}
    and from \cref{geodesic-distance-LR},
    \begin{align}
    	S_{AB}=2\times\frac{1}{2G_N}\log\left[\frac{r_\infty}{\sqrt{L}}\left(1+\frac{\alpha}{2}\right)\right]\,.
    \end{align}
    For a lage black hole ($M\gg 1$), one obtains the mutual information to be
    \begin{align}
    	I(A:B)\sim\frac{1}{G_N}\left[\log\sinh\frac{\pi\ell}{\sqrt{\beta^2-8\pi^2\mu}}-\lambda_L t_W-\log\left(\frac{E}{8M}\frac{1+2\mu\CL_\mu}{1-2\mu\CL_\mu}\right)\right]\,.\label{MI}
    \end{align}
    \paragraph{Scrambling time:} The scrambling time is defined as the timescale at which the mutual information vanishes, $I(A:B)=0$. In our case, the scrambling time may be obtained from \cref{MI} as follows
    \begin{align} t_\star&=\frac{\beta}{2\pi}\log\left[\frac{8M}{E}\sqrt{1-\frac{8\pi^2\mu}{\beta^2}}\sinh\frac{\pi\ell}{\sqrt{\beta^2-8\pi^2\mu}}\right]\,,\notag\\
    	&\sim \frac{\ell}{2\sqrt{\beta^2-8\pi^2\mu}}+\frac{\beta}{2\pi}\log\left[\frac{8M}{E}\sqrt{1-\frac{8\pi^2\mu}{\beta^2}}\right]\,.
    \end{align}
    
%	\section*{Acknowledgment}
%	These notes are based on preliminary derivations and serve as a starting point for more detailed analysis.
	%----------------------------------------------------------------------------
	\bibliographystyle{JHEP}
	\bibliography{reference}
\end{document}